# Magnetic fields do not suppress global star formation in low metallicity dwarf galaxies

David J. Whitworth,[1,2]★ Rowan J. Smith,[2] Ralf S. Klessen,[1,3] Mordecai-Mark Mac Low,[6] Simon C. O. Glover,[1] Robin Tress,[4] Rüdiger Pakmor,[5] Juan D. Soler[7]

[1] *Universitat Heidelberg, Zentrum fur Astronomie, Institut fur Theoretische Astrophysik, Albert-Ueberle-Str. 2, 69120 Heidelberg, Germany*
[2] *Jodrell Bank Centre for Astrophysics, Department of Physics and Astronomy, University of Manchester, Oxford Road, Manchester M13 9PL, UK*
[3] *Universität Heidelberg, Interdisziplinäres Zentrum für Wissenschaftliches Rechnen, Im Neuenheimer Feld 205, D-69120 Heidelberg, Germany*
[4] *Institute of Physics, Laboratory for galaxy evolution and spectral modelling, EPFL, Observatoire de Sauverny, Chemin Pegais 51, 1290 Versoix, Switzerland.*
[5] *Max-Planck-Institut für Astrophysik, Karl-Schwarzschild-Str. 1, 85741 Garching, Germany*
[6] *Department of Astrophysics, American Museum of Natural History, 79th St. at Central Park West, New York, NY 10024, USA*
[7] *Istituto di Astrofisica e Planetologia Spaziali (IAPS). INAF. Via Fosso del Cavaliere 100, 00133 Roma, Italy*

**ABSTRACT**

Many studies concluded that magnetic fields suppress star formation in molecular clouds and Milky Way like galaxies. However, most of these studies are based on fully developed fields that have reached the saturation level, with little work on investigating how an initial weak primordial field affects star formation in low metallicity environments. In this paper, we investigate the impact of a weak initial field on low metallicity dwarf galaxies. We perform high-resolution AREPO simulations of five isolated dwarf galaxies. Two models are hydrodynamical, two start with a primordial magnetic field of $10^{-6}\mu$G and different sub-solar metallicities, and one starts with a saturated field of $10^{-2}\mu$G. All models include a non-equilibrium, time-dependent chemical network that includes the effects of gas shielding from the ambient ultraviolet field. Sink particles form directly from the gravitational collapse of gas and are treated as star-forming clumps that can accrete gas. We vary the ambient uniform far ultraviolet field, and cosmic ray ionization rate between 1% and 10% of solar values. We find that the magnetic field has little impact on the global star formation rate, which is in tension with some previously published results. We further find that the initial field strength has little impact on the global star formation rate. We show that an increase in the mass fractions of both molecular hydrogen and cold gas, along with changes in the perpendicular gas velocity dispersion and the magnetic field acting in the weak-field model, overcome the expected suppression in star formation.

## 1 INTRODUCTION

The impact of the magnetic field **B** on the evolution of galaxies and on star formation has long been debated. However, as the magnetic energy and turbulent energy densities are observed to be in equipartition in nearby galaxies (Beck et al. 1996), magnetic fields must be dynamically important in galaxies.

The Milky Way and similar galaxies are often used as test beds to study the effects of magnetic fields on star formation. We know that the Milky Way has a galactic field strength of ∼ 2–10 μG (Beck 2001; Brown et al. 2003), and in the Solar neighbourhood it is estimated at 6 μG (Beck 2001). But is the Milky Way a good representation of other galaxies and environments? Are the field strengths the same for example in low metallicity regions where cooling may be weaker, the star formation rate (SFR) is lower and there is stronger stirring of the interstellar medium (ISM) by supernova (SN) feedback?

The magnetic field is dynamically important in regions where the magnetic pressure is similar to or exceeds thermal or turbulent pressure, whilst elsewhere the field is carried along by the flow in the kinematic regime. Hennebelle & Inutsuka (2019) and Krumholz & Federrath (2019) have shown that a dynamically important magnetic field can delay star formation and suppress the SFR in clouds and isolated Milky Way like galaxies by up to a factor of two due to the field slowing the collapse of molecular clouds. At cloud scales, models have shown that the SFR can be suppressed by 1–2 orders of magnitude (Price & Bate 2009; Federrath & Klessen 2013; Zamora-Avilés et al. 2018), delayed or completely suppressed (Körtgen & Banerjee 2015) by a sufficiently strong B-field. At larger scales, magnetized turbulence dominates over gravity (Ibáñez-Mejía et al. 2022; Pattle et al. 2022), which can shape the ISM, reducing the number of cores and clumps (Hennebelle & Inutsuka 2019).

Cloud-scale models have also shown that the formation of dense molecular gas is delayed by as much as ∼ 25 Myr when a B-field is included (Girichidis et al. 2018). If star formation is connected to molecular gas, this would delay its onset of star formation. In models with varying field strengths, low strengths have been shown to increase the mass fraction of molecular gas (Pardi et al. 2017).

★ E-mail: dd163@uni-heidelberg.de

Observations of magnetic fields, revealed by the dust polarized thermal emission and the Zeeman effect toward nearby molecular clouds (MCs), indicate that magnetization plays an important role in structuring the ISM on parsec scales and densities up to $10^3$ cm$^{-3}$ (see Crutcher 2012; Pattle & Fissel 2019, for recent reviews). Diffuse gas structures traced by atomic hydrogen emission appear preferentially aligned with the local magnetic field (Clark et al. 2014). Toward MCs, the relative orientation between the column density and the magnetic field changes from preferentially parallel to preferentially perpendicular with increasing column density (Planck Collaboration et al. 2016; Heyer et al. 2020). This change in relative orientation has been interpreted as the combined effect of the anisotropy, caused by a relatively strong magnetic field, and a converging flow, such as that produced by gravitational collapse (Soler & Hennebelle 2017; Seifried et al. 2020; Barreto-Mota et al. 2021; Ibáñez-Mejía et al. 2022). It can be interpreted as the magnetic field becoming sub-dominant in the cloud dynamics, and thus being unable to reduce the star formation rate, as has also been suggested by observations (Soler 2019). However, the magnetic field may still be significant for the overall star formation by setting the amount of molecular gas available to form new stars (see, for example, Ntormousi et al. 2017; Girichidis et al. 2021; Kim et al. 2021).

Most theoretical models of magnetized galaxies start from a saturated magnetic field. In reality the field must be amplified from a primordial seed field by a dynamo. The origin of the primordial magnetic field is still much debated (Kulsrud & Zweibel 2008; Mandal et al. 2022), but it is expected to have a magnitude around $10^{-16}$ G (Neronov & Vovk 2010; Dolag et al. 2011). This is many orders of magnitude below what has been observed within the Milky Way. Within the context of galaxy evolution there are two main processes that amplify the B-field, the small-scale or fluctuating dynamo (SSD) and the large-scale or mean-field dynamo (LSD) (Brandenburg & Subramanian 2005).

A number of numerical studies have shown the importance of the SSD on field amplification within the ISM (Sur et al. 2010; Federrath et al. 2011; Schober et al. 2012; Rieder & Teyssier 2016, 2017; Gent et al. 2021). It is, however, difficult to study such small-scale effects in galactic models due to resolution and computational limits. High resolution cloud scale models have shown that the SSD is important for driving structure formation, as it can suppress gas fragmentation and is especially important in the primordial gas that formed the first stars (Federrath et al. 2014; Sharda et al. 2021). The SSD around the first stars is most likely powered by astrophysical turbulence driven by accretion from the local environment due to gravity driven flows at multiple scales (Klessen & Hennebelle 2010; Elmegreen & Burkert 2010). In star-forming galaxies, however, turbulence driven by supernovae also becomes important in driving SSDs (e.g. Balsara & Kim 2005; Steinwandel et al. 2020).

Pakmor et al. (2017) argue that amplification of a primordial field in a cosmological zoom-in of Milky Way sized galaxies can increase or decrease the star formation rate and that the impact of magnetic fields on the dynamical evolution of a galaxy is minimal due to reaching equipartition too late to affect most star formation. However, it has been demonstrated (Balsara et al. 2004; Sur et al. 2010; Gent et al. 2021) that the speed of amplification depends critically on numerical resolution, with convergence still not being reached in local models with sub-parsec resolution, so fields may grow far more promptly than expected in even zoom-in simulations.

It has also been shown in similar zoom-in models (Martin-Alvarez et al. 2020) that a strong primordial field of around $10^{-2}$ $\mu$G can delay the onset of star formation and shrink the star-forming disc by up to a factor of 1/3 as the field takes energy from the rotational support, resulting in more gas in the centre of the galaxy.

All these studies have concentrated on Milky Way like galaxies, though it is thought that dwarf galaxies are the most numerous galaxies in the universe and likely the progenitors of larger galaxies through mergers. Having an understanding of their magnetic fields and how they affect star formation is key to understanding the evolution of larger galaxies and their B-fields. Local Group dwarf galaxies show an average field strength of $< 4.2 \pm 1.8$ $\mu$G (Chyży et al. 2011), a bit lower than the Milky Way average. This implies that there maybe a link between SFR and B-field strength.Little work has been done on understanding how the amplification of primordial fields to saturation affects star formation in low-metallicity dwarf galaxies. We aim to address this here. This paper is organised as follows: In Section 2 we lay out the numerical models used including the application of the MHD code. In Section 3 we detail the results from the new simulations. Section 4 discusses the implications of the results in comparison to previous studies and the caveats. In Section 5 we provide a summary of our key results.

## 2 METHODS

### 2.1 Numerical Modelling

We use the moving mesh code AREPO (Springel 2010). This solves the ideal magnetohydrodynamic (MHD) equations using an unstructured Voronoi mesh that allows the cells to move with the local gas velocity (Pakmor et al. 2011). The mesh is restructured after each time step, allowing for a continuous and adaptive resolution while allowing the mass in the cells to vary by up to a factor of two. In order to avoid artificial fragmentation and ensure collapsing gas is fully resolved, the Jeans (1902) length is always resolved by at least 8 cells (Truelove et al. 1997; Federrath et al. 2011). To mitigate the divergence errors that arise in MHD models this scheme implements the Powell et al. (1999) divergence cleaning method along with an HLLD Riemann solver (Miyoshi & Kusano 2005). In order to consistently compare the models we use the same solver for both the hydro and MHD cases.

To model the ISM we use a non-equilibrium, time-dependent, chemical network based on the work of Gong et al. (2017) including the effects of cosmic-ray ionization. We model self-shielding of dissociating ultraviolet (UV) radiation using the TreeCol algorithm of Clark et al. (2012) and use a shielding radius of 30 pc. The molecular and atomic cooling functions are the same as those in Clark et al. (2019), as originally described by Glover et al. (2010). The network allows us to directly trace nine non-equilibrium species ($H_2$, $H^+$, $C^+$, $CH_x$, $OH_x$, $CO$, $HCO^+$, $He^+$ and $Si^+$) and derive a further eight from conservation laws (free electrons, H, $H_3^+$, C, O, $O^+$ He and Si). Full details of how our chemical network differs from that presented in Gong et al. (2017) can be found in Hunter et al. (2021). We do not include photoionization feedback in these models due to computational constraints.

To model star formation we use an accreting sink particle model. To ensure each sink represents a region of gas that is collapsing and truly self gravitating, the gas above a threshold density, $\rho_c$, of $2.4 \times 10^{-21}$ g cm$^{-3}$ (equivalent to $n = 10^3$ cm$^{-3}$) and within an accretion radius, $r_{acc}$, of 1.75 pc must satisfy the following criteria:

(i) The gas flow must be converging with negative divergence of both velocity and acceleration.

(ii) The region must be centred on a local potential minimum.

(iii) The region must not fall within the accretion radius of another sink particle.

(iv) The region should not move within the accretion radius of other sink particles in a time less that the local free-fall time.

(v) The region must be gravitationally bound.

Once a sink is formed, it will accrete gas from the surrounding ISM if the gas is within $r_{acc}$, above $\rho_c$ and gravitational bound to the sink. The accreted mass is limited to 90% of the initial mass of the sink per timestep and we only skim mass from the cell up to the density threshold so we do not allow runaway growth of the sink. The mass in the sink is then associated with a discrete stellar population using the method from Sormani et al. (2017). We apply an assumed star formation efficiency in the sink of $\epsilon_{SF} = 0.1$, with the remaining 90% of the mass considered to be gas. Each star more massive than $8 M_\odot$ that reaches the end of its lifetime will generate a supernova explosion (SNe).

The SNe inject energy and momentum back into the ISM using the method described in Whitworth et al. (2022, hereafter Paper 1) and Tress et al. (2020). The injection radius $R_{inj}$ of 16 cells is compared to the radius of the remnant at the end of its Sedov-Taylor phase ($R_{ST}$). If $R_{ST} > R_{inj}$ then we inject $10^{51}$ erg into the gas as thermal energy and fully ionise the gas. If the opposite is true, i.e., the local gas density is too high for thermal injection, we directly inject momentum into the ISM, setting the temperature in the bubble to $10^4$ K and allow the ionisation state of the gas to evolve from the injected temperature and not artificially changing it.

Each SNe also injects mass, $M_{in}$, into the ISM via $M_{in} = (M_{sink} - M_{stars})/nSN$, where $M_{sink}$ is the mass of the sink at the time of the SNe, $M_{stars}$ is the mass of stars in the sink and nSN is the number of remaining SNe that the sink will produce. This mass is uniformly distributed across cells in the energy injection region.

Once the last SN from a sink particle has occurred, we consider the mass remaining in the sink to be only the stellar mass. At this point the sink particle is converted into a collisionless $N$-body star particle. If a sink has no associated SNe due to being populated by low mass stars, then after 10 Myrs a fraction $\epsilon_{SF}$ of its mass is turned into a star particle and the remaining gas is returned to the surrounding gas cells.

For a more detailed description of the code and our custom physics modules, see Paper 1.

## 2.2 Initial Conditions

We model an isolated dwarf galaxy as a stable disc that is made up of two components: a dark matter halo and a gaseous disc. In the initial conditions we do not include star particles, but allow them to form in an initial burst of star formation that occurs before the galaxy settles into the steady-state where we will perform our analysis.

The dark matter halo follows a spheroidal Hernquist (1990) density profile:

$$\rho_{sph}(r) = \frac{M_{sph}}{2\pi} \frac{a}{r(r+a)^3},$$  (1)

where $r$ is the spherical radius, $a = 7.62$ kpc is the scale-length of the halo and $M_{sph} = 2 \times 10^{10} M_\odot$ is the mass of the halo. The gaseous disc component follows a double exponential density profile:

$$\rho_{disc}(R, z) = \frac{M_{disc}}{2\pi h_z h_R^2} \text{sech}^2 \left( \frac{z}{2h_z} \right) \exp \left( -\frac{R}{h_R} \right),$$  (2)

where $R$ is the cylindrical radius and $z$ is height above the midplane, $M_{disc} = 8 \times 10^7 M_\odot$ is the mass of the gas in the disc, and $h_z =$

| | Mass ($M_\odot$) | Scale length (kpc) | $h_z$ (kpc) |
|---|---|---|---|
| DM Halo | $2.00 \times 10^{10}$ | 7.62 | - |
| Gas disc | $8.00 \times 10^7$ | 0.82 | 0.35 |

**Table 1.** Parameters for the different galactic components

| Model | Z ($Z_\odot$) | $\log_{10} \xi$ | $\zeta_H$ (s$^{-1}$) | $G$ ($G_0$) | $\log_{10} B_0$ ($\mu$G) |
|---|---|---|---|---|---|
| HYD_10 | 0.10 | -3 | $3 \times 10^{-18}$ | 0.17 | 0 |
| HYD_01 | 0.01 | -4 | $3 \times 10^{-19}$ | 0.017 | 0 |
| MHD_10 | 0.10 | -3 | $3 \times 10^{-18}$ | 0.17 | -6 |
| MHD_01 | 0.01 | -4 | $3 \times 10^{-19}$ | 0.017 | -6 |
| MHD_SAT | 0.10 | -3 | $3 \times 10^{-18}$ | 0.17 | -2 |

**Table 2.** Values used in the initial conditions for each model for metallicity $Z$, dust-to-gas ratio $\xi$, relative to the value in solar metallicity gas, cosmic ionisation rate $\zeta_H$, UV field strength $G$, and magnetic field strength $B_0$.

0.35 kpc and $h_R = 0.82$ kpc are its scale-height and scale-length, respectively.

Our initial conditions are generated using the method from Springel (2005) with values chosen to be broadly comparable to suggestions by Hu et al. (2016). Table 1 shows the galactic components and their values. Table 2 shows the values used for metallicity $Z$, dust-to-gas ratio $\xi$ (which has a Solar neighborhood value of $\xi = 0.01$), cosmic ray ionisation rate $\zeta_H$, far-ultraviolet (FUV) field strength $G$ given in Habing units $G_0$, and the MHD seed field strength $B_0$. We vary $Z_\odot$, $\xi$, and $G$ from 1–10% of Solar values. We take the fiducial hydrodynamical, model HYD_10, run from our previous paper (Whitworth et al. 2022, hereby referred to as Paper 1), and the 1% metallicity 1% UV-field for model HYD_01. For a more detailed look at the effects of the UV field on dwarf galaxies and the ISM, we refer the reader to Hu et al. (2016, 2017).

For the MHD models we apply a uniform, mono-directional, poloidal seed field of $10^{-6} \mu$G in models MHD_01 and MHD_10, while in model MHD_SAT we apply a field of $10^{-2} \mu$G to represent a saturated field. Our choice of field strength is inspired by Pakmor et al. (2014), which is higher than the lower limit of $10^{-16}$ G for the primordial field (Neronov & Vovk 2010; Taylor et al. 2011). We start higher than this for computational expediency, though being higher does not matter as the field amplifies quickly from primordial strengths via the SSD and would reach our level in a short time frame, with the LSD only becoming important at later times. The initial orientation of the field is poloidal, though this is unimportant as the galactic dynamo will redirect the field to follow the rotation of the disc in a few Myr (Pakmor et al. 2014).

The initial temperature of the gas is set to $T=10^4$ K and the chemical composition is initially considered to be fully atomic.

## 2.3 Resolution

For all models we set a base target mass resolution for the gas cells to 50 M$_\odot$. The default refinement scheme in AREPO refines and de-refines cells such that the mass stays within a factor of 2 of our target value. On top of this we also apply a Jeans refinement criterion so that the Jeans Length is resolved by at least 8 resolution elements. This means that the mass per cell is substantially smaller than 50 M$_\odot$ at high densities, $\sim 0.1$ M$_\odot$ at $n = 10^3$ cm$^{-3}$. In the MHD models, at a number density of $n = 100 \, \text{cm}^{-3}$ we have an average cell radius ($r_{\text{cell}}$) of 0.79 pc, and at an $n = 10^3 \, \text{cm}^3$ the $r_{\text{cell}}$ is 0.31 pc . In order to avoid artificial fragmentation and make sure we are resolved in collapsing gas we ensure the Jeans length is always resolved by at least 8 cells (Jeans 1902; Truelove et al. 1997; Federrath et al. 2011).

As the cells range in size and mass, we use adaptive softening for the gas cell size, and apply a fixed softening length to the dark matter of 64 pc and to the sink particles of 2 pc.

## 3 RESULTS

In this section we begin by examining the difference in the morphology of the models, looking at general structure and gas distribution. Next we look at how the field grows and evolves over the time of the simulation. We then look at the effects a magnetic field has on the basic ISM properties in the galaxy and the impact of magnetic pressure in terms of plasma-$\beta$, equation 5.

Our comparisons are made over a steady state period that we define from 300 Myr to 1 Gyr. This allows enough time for both the star formation and chemical evolution to have settled (see section 3.4). An argument could be made to start the steady state period later in models MHD_01 and MHD_10 as the B-field is still growing at this point, but as we shall see, this has little effect on star formation and chemical evolution.

### 3.1 Morphology

All models are run for 1 Gyr. In all cases we see an initial burst of star formation that creates a large number of SNe that clear the gas away from the central region of the galaxy. This leads to the formation of a ring-like structure which then fills in slowly over time. The remnants of the burst can be seen in the 500 Myr snapshot for models MHD_10, MHD_SAT and HYD_10 in Figure 1. For models MHD_01 and HYD_01 this structure has already been lost and a steady state morphology has been reached.

Figure 1 shows the face-on distribution of $\Sigma_{\text{HI}}$ and $\Sigma_{\text{H}_2}$, excluding the gas locked in sink particles. At comparable metallicities the radial extent of the $\text{H}_2$ is similar in both the MHD and HD cases. However, more of the disc surface contains $\text{H}_2$ in the MHD case, i.e. there is a greater filling factor of $\text{H}_2$, calculated as the percentage of gas above $\Sigma_{\text{H}_2} = 10^{-5} \, \text{M}_\odot \, \text{pc}^{-2}$ that covers a circle of radius $r = 2.5$ kpc in each plot of Figure 1, Table 3. Considering the edge-on distribution in Figure 2 and looking at the vertical distribution, $z$ Figure 3, we can see that the $\Sigma_{\text{H}_2}$ for MHD_01 has a greater distribution in $z$ than the corresponding HD model. MHD_10 and MHD_SAT show a similar distribution in $z$ to HYD_10. At $\Sigma_{\text{H}_2} = 10^{-1} \, \text{M}_\odot \, \text{pc}^{-2}$ the height of the discs are: MHD_01 = 0.34 kpc, HYD_01 = 0.22 kpc, MHD_10 = 0.26 kpc, HYD_10 = 0.26 kpc, MHD_SAT = 0.21 kpc. The $\Sigma_{\text{HI}}$ for all models shows no variation in scale height. From this we define a star forming disc, $r = 2.5$ kpc, $|z| < 0.4$kpc, this is the volume of the disc where all star formation takes place.

### 3.2 Magnetic Field Growth

Figure 4 shows the growth of the mass-weighted B-field over the full 1 Gyr of the simulations within the dense ($n > 1 \, \text{cm}^{-3}$) star forming disc. To show that our choice of initial field direction is not important we plot the growth of the B-field for the poloidal (dashed line) and toroidal (dotted line) components and the total components (solid line). From this we can see that the toroidal component becomes dominant over the poloidal component very quickly, after roughly

| Model | 500 Myr | 750 Myr | 1 Gyr |
|---|---|---|---|
| MHD_01 | 6.66% | 8.10% | 10.82% |
| HYD_01 | 4.60% | 3.76% | 5.46% |
| MHD_10 | 3.16% | 6.64% | 7.94% |
| HYD_10 | 2.96% | 4.67% | 4.18% |
| MHD_SAT | 6.01% | 7.17% | 7.80% |

**Table 3.** Percentage (Filling Factor) of the projection of the star forming disc ($r = 2.5$ kpc) covered by $\text{H}_2$ above $\Sigma_{\text{H}_2} = 10^{-5} \, \text{M}_\odot \, \text{pc}^{-2}$ from Figure 1. We can see that the MHD models have a larger filling factor than the HD models.

20 Myr for all models. The B-field grows from the initial seed fields for all models and the MHD_01 and MHD_10 runs saturate at a level of $\sim 5\mu$G and MHD_SAT at $\sim 10\mu$G. Model MHD_SAT reaches saturation quickly, after $\sim 200$ Myr, before the beginning of the steady state period. Models MHD_01 and MHD_10 take longer to reach saturation due to starting with much lower initial seed field strengths. To see if this is due to resolution or rotation period we run a lower resolution model and include the results in Appendix A where we see that the B-field grows more slowly at lower resolution but that this only has a minimal effect on the star formation rate and the depletion time of the gas. MHD_01 shows a steady growth for $\sim 400$ Myr and then a slow down in growth until it reaches saturation at $\sim 600$ Myr. MHD_10 has an initial burst in B-field growth during the initial starburst, with a small decrease after the burst ends. It then has a period of more varied growth, with periods of growth and slight decline until it reaches saturation at $\sim 700$ Myr.

Figure 5 shows a projection of the B-field across the disc, and in the $z$-plane through the disc at 1 Gyr. The stream lines show the strength and orientation of the field lines. In models MHD_01 and MHD_10 where the B-field is self-consistently generated, even after 1 Gyr it appears disordered within the disc. When a higher "saturated" B-field is used as an initial condition, the field becomes ordered by 1 Gyr with the field lines following the rotation of the disc.

Figure 6 shows the molecular gas mass-weighted absolute B-field in $\mu$G against number density ($n$) in $\text{cm}^{-3}$ for each model at 1 Gyr. MHD_SAT shows little $|B|$ below $10^{-2} \, \mu$G as this was the initial strength of the B-field. All three models show an increasing B-field strength as number densities increase. The slopes shown with the coloured dashed lines indicate how $|B|$ increases with density for field strengths $> 10^{-3} \, \mu$G. We find that MHD_01 has a slope of 0.55, MHD_10 = 0.58, and MHD_SAT = 0.46. We make this cut so as to avoid the low density gas at the edge of the star forming disc while still sampling the weak $|B|$ and low density regime. All models show that most of the dense molecular gas, n > 10 $\text{cm}^{-3}$, exists in regions with a $|\bar{B}|$ of greater than 1 $\mu$G.

### 3.3 Star Formation

To calculate the star formation rate (SFR) we take the stellar mass accreted ($M_{\odot,\text{acc}}$) onto the sink particles at each snapshot, add the stellar mass from new sinks created ($M_{\odot,\text{new}}$) in the timestep, and then divide by the time difference between the current snapshot ($t_2$) and the previous snapshot ($t_1$) ($\sim 9.78 \times 10^5$ Myr):

$$SFR = \frac{(M_{\odot,\text{acc}} + M_{\odot,\text{new}})}{t_2 - t_1} . \tag{3}$$

In Figure 7 we show a comparison of the star formation rates (SFR) between all models over the steady state period. From this data there is no large variation between the MHD and hydrodynamical models. There is a small but noticeable increase in the SFR from HYD_01 by dividing the SFR$_{ss}$ by the average gas mass of the star-forming disc during the steady state period. It shows no significant variation across all models.

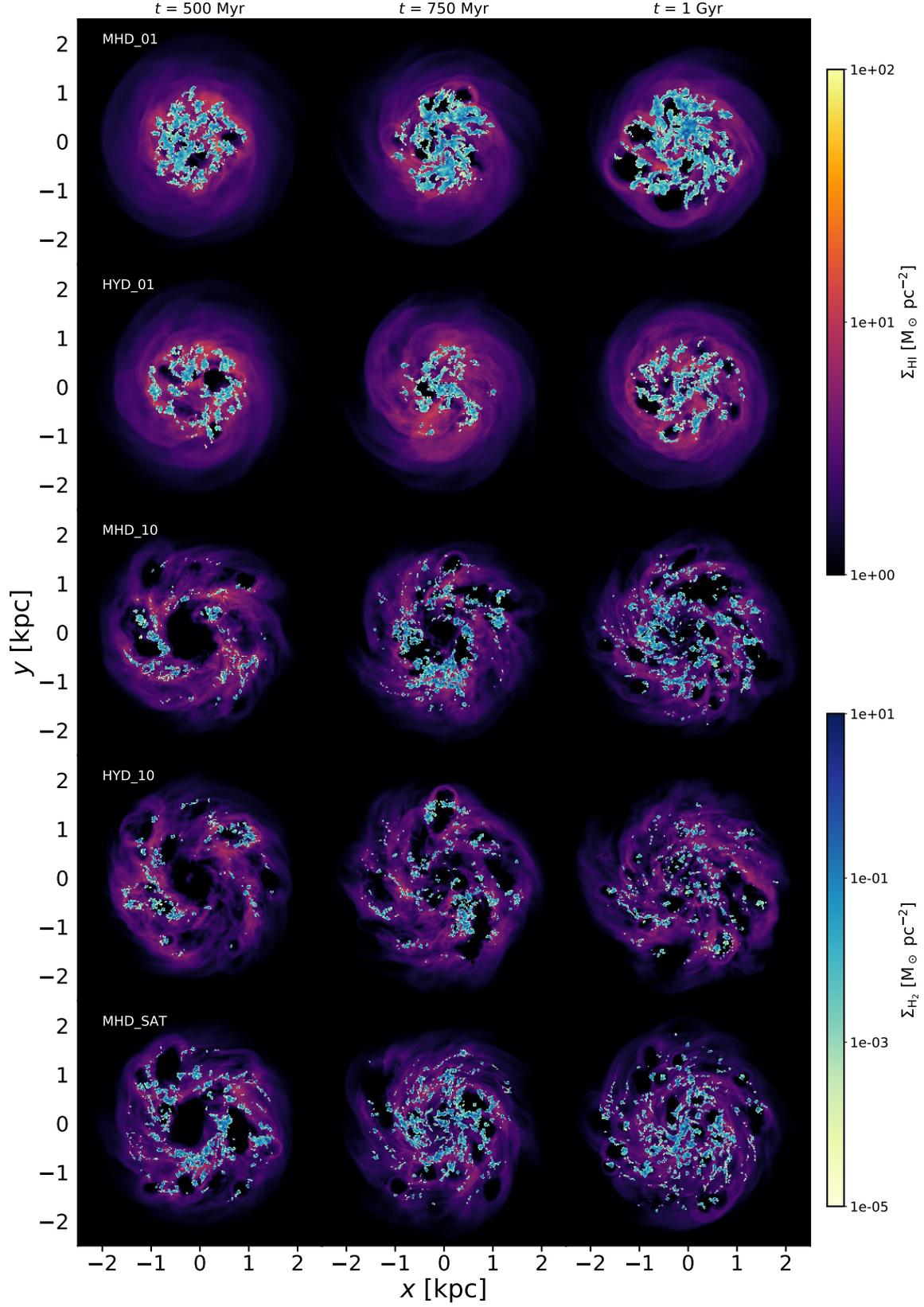

**Figure 1.** The HI surface density ($\Sigma_{HI}$) and H$_2$ surface density ($\Sigma_{H_2}$) for all models in the $xy$ plane. From left to right, we show the disc at times $t$ = 500 Myr, 750 Myr and 1000 Myr. From the top going down we plot MHD_01, HYD_01, MHD_10, HYD_10, MHD_SAT. It can be seen that the $\Sigma_{H_2}$ in all MHD models is greater when compared to the corresponding hydrodynamical models. Looking at MHD_01 we see that the distribution of $\Sigma_{H_2}$ is much greater than in HYD_01. In MHD_10 and MHD_SAT there is only slightly more $\Sigma_{H_2}$ than in HYD_10 with a similar distribution.

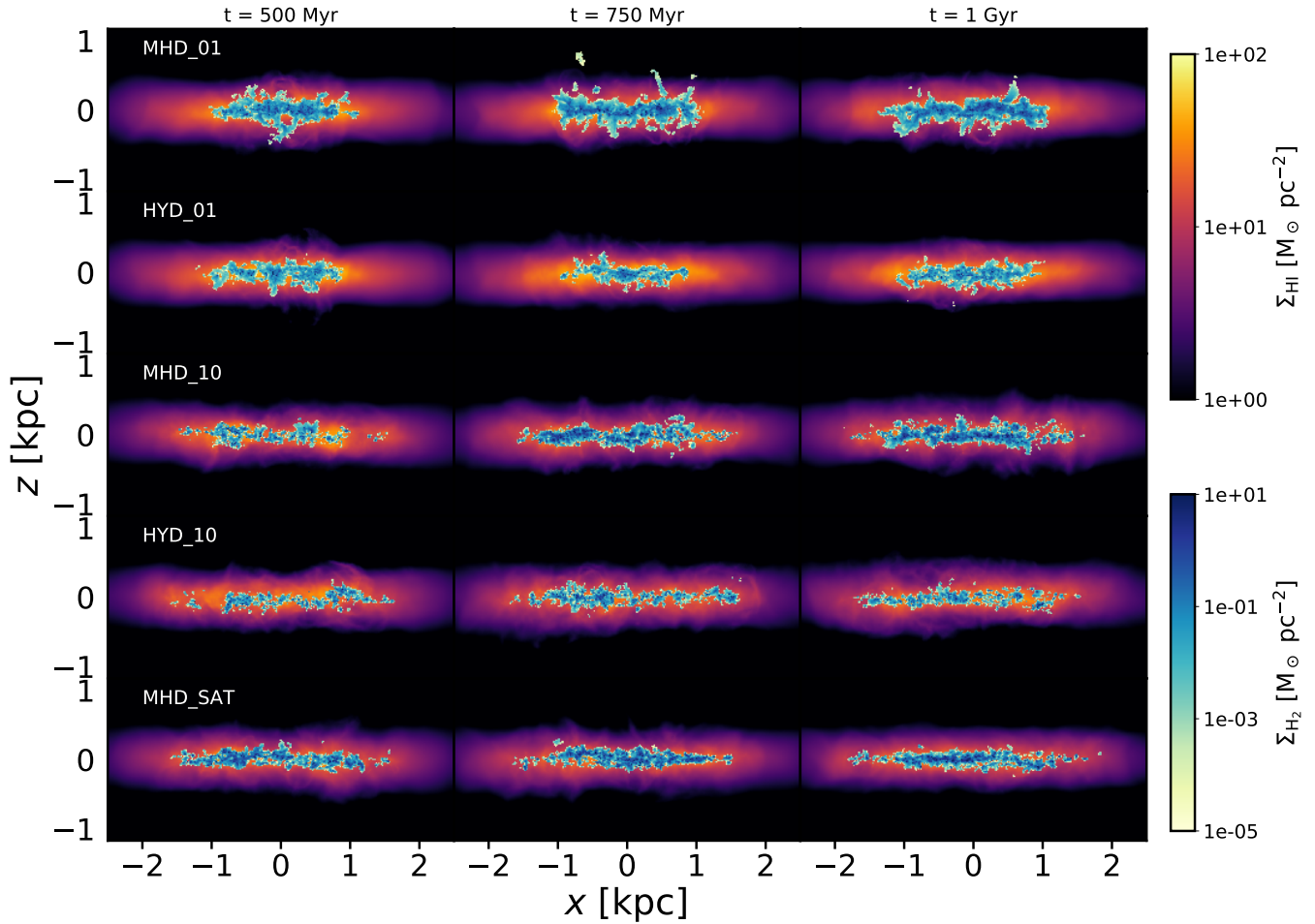

**Figure 2.** The HI surface density ($\Sigma_{HI}$) and $H_2$ surface density ($\Sigma_{H_2}$) for all models in the $xz$ plane. Plotted as in Figure 1. The $\Sigma_{H_2}$ has a greater scale height in the MHD_01 and MHD_10 models with a greater surface density over all compared to HYD_01 and HYD_10 models respectively. In the MHD_SAT model the $\Sigma_{H_2}$ is more confined to the disc with a similar scale height compared to the HYD_10 model, but with a greater distribution.

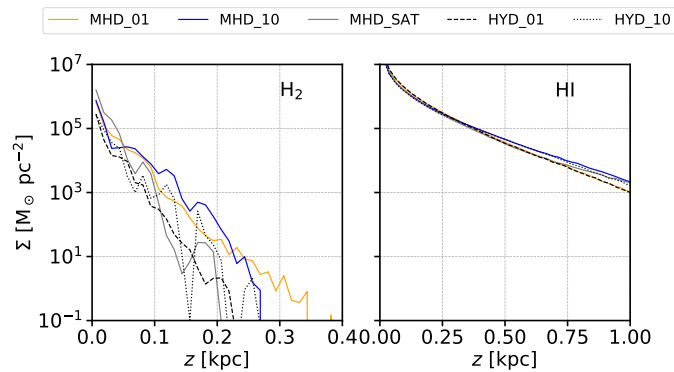

**Figure 3.** HI surface density ($\Sigma_{HI}$) and $H_2$ surface density ($\Sigma_{H_2}$) as a function of height above the centre of the plane of the disc for all models. The MHD models show a greater distribution in $H_2$ scale height than the corresponding HD models. There is little difference in $\Sigma_{HI}$.

| | SFR$_{ss}$ (M$_\odot$ yr$^{-1}$) | $\sigma_{ss}$ | sSFR (yr$^{-1}$) |
|---|---|---|---|
| MHD_01 | $2.77 \times 10^{-3}$ | $1.19 \times 10^{-3}$ | $3.46 \times 10^{-11}$ |
| HYD_01 | $1.90 \times 10^{-3}$ | $0.82 \times 10^{-3}$ | $2.4 \times 10^{-11}$ |
| MHD_10 | $1.84 \times 10^{-3}$ | $0.92 \times 10^{-3}$ | $2.20 \times 10^{-11}$ |
| HYD_10 | $2.01 \times 10^{-3}$ | $0.96 \times 10^{-3}$ | $2.5 \times 10^{-11}$ |
| MHD_SAT | $1.47 \times 10^{-3}$ | $0.48 \times 10^{-3}$ | $1.83 \times 10^{-11}$ |

**Table 4.** The steady state star formation rates (SFR$_{ss}$), their standard deviations and specific SFR (star formation per unit gas mass, sSFR) of the two models.

### 3.4 ISM Chemistry

Figure 8 shows the $H_2$ mass-weighted (top row) and total gas weighted (bottom row) phase diagram (temperature against number density) for models MHD_10 (left) and HYD_10 (right) excluding to MHD_01, and a small decrease from HYD_10 to MHD_10 and MHD_SAT, but these changes are not substantial. MHD_SAT shows a reduction in stochasticity and has no large scale variability compared to MHD_10. Table 4 shows the steady-state SFR (SFR$_{ss}$) for each model and the corresponding specific SFR (sSFR), calculated the gas trapped in the sinks [1]. The plots show an increase in the mass of $H_2$ in the cold phase for MHD_10 when compared to HYD_10. In addition we see more $H_2$ in general in MHD_10, with a large proportion in the warm phase. We note that the warm phase is dominant in all our models but focus on the cold phase here as our interest is in the star forming gas.

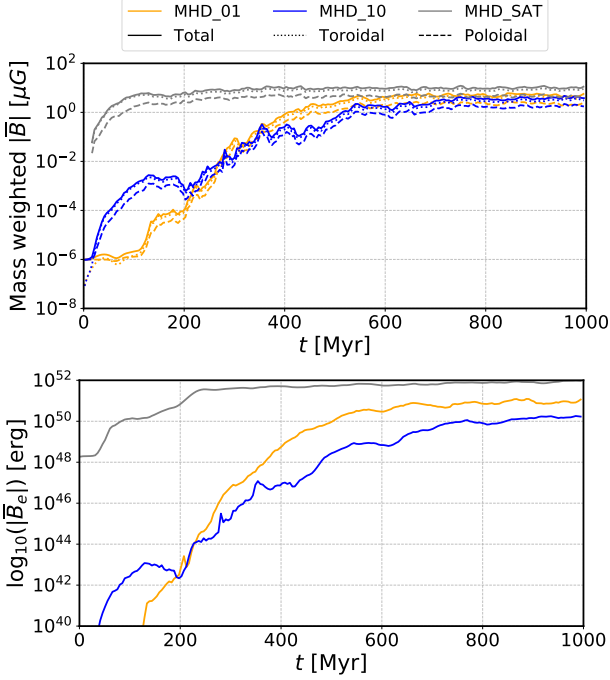

**Figure 4.** Top panel: The growth of the mass-weighted mean B-field ($|\overline{B}|$) in $\mu$G in the dense ($n > 1\mathrm{cm}^{-3}$) star forming disc over the full simulation. We plot the growth for different components of the B-field for each model, MHD_01 in orange, MHD_10 in blue and MHD_SAT in grey. The solid line represents the total B-field, the dashed line represents the poloidal component from initial conditions and the dotted line is the toroidal component that forms as the field evolves. All models reach saturation at different times, with the MHD_01 and MHD_10 saturating at a level of $\sim 5\mu$G and MHD_SAT at $\sim 10\mu$G. Bottom panel: Total magnetic energy $|\overline{B}_{e}|$ of the star forming disc over the simulation.

We plot in Figure 9 the mass fractions for molecular hydrogen and cold gas ($T \leqslant 100$ K) as fractions of the total mass. These should be considered lower limits as we do not include mass from the sink particles. The MHD models have higher mass fractions for both $H_2$ and cold gas when compared to the corresponding hydrodynamical models. At the start of the steady state period, MHD_01 and MHD_10 have similar mass fractions compared to the hydrodynamical cases, but these then increase over the steady state period as the field amplifies. Table 5 shows the percentage mass fraction for each model. The mass fraction increases by a factor 3 for $H_2$ in model MHD_01, and a factor 2 in the cold gas in the same model. MHD_SAT shows a factor of 2 increase in both mass fractions whilst MHD_10 shows an increase by a factor of 1.6 in $H_2$ and 1.3 in cold gas.

Figure 10 compares the gas depletion times ($\tau_{\rm dep}$) for $H_2$ (top) and cold gas (bottom) for each model:

$$\tau_{\rm dep}(\rm gas) = \frac{M_{\rm gas}}{SFR},\qquad(4)$$

Table 5 gives the average steady state $\tau_{\rm dep}$ for each model. Looking

---

| model | $\frac{M_{\rm H_2}}{M_{\rm HI+H_2}}$ | $\frac{M_{\rm sold}}{M_{\rm HI+H_2}}$ | H2 | | cold gas | |
|---|---|---|---|---|---|---|
| | | | $\tau_{\rm dep}$ (Myr) | $\sigma_{\tau_{\rm dep}}$ (Myr) | $\tau_{\rm dep}$ (Myr) | $\sigma_{\tau_{\rm dep}}$ (Myr) |
| MHD_01 | 0.24% | 1.17% | 73.3 | 36.5 | 355 | 173 |
| HYD_01 | 0.08% | 0.58% | 35.4 | 17.3 | 257 | 116 |
| MHD_10 | 0.43% | 2.20% | 163 | 61.6 | 835 | 312 |
| HYD_10 | 0.27% | 1.66% | 91.6 | 26.3 | 566 | 165 |
| MHD_SAT | 0.56% | 3.24% | 263 | 83.2 | 1540 | 471 |

**Table 5.** Mass fractions and depletion times $\tau_{\rm dep}$ for $H_2$ and cold HI + $H_2$ gas with their standard deviations $\sigma_{\tau_{\rm dep}}$. The MHD models show a higher mass fraction in both $H_2$ and cold gas compared to the corresponding HD models, by a factor of 2 or higher. The $\tau_{\rm H_2\ dep}$ and $\tau_{\rm cold\ dep}$ for the MHD models are longer than for the HD models. As there is more gas mass, therefore a larger reservoir to deplete, resulting in longer depletion times.

at the $\tau_{\rm dep}$ in all models for both gas components, we see the MHD models have longer times on average. In MHD_01 and MHD_10 the depletion times at the start of the steady state are similar to HYD_01 and HYD_10 respectively, but grow longer over the steady state period. This matches the growth in the mass fractions. As the reservoir of gas grows so does the depletion time. MHD_SAT has a roughly constant depletion time for both gas components. As the mass fractions have increased over the same period, there is a larger reservoir to deplete, hence an increase in depletion times.

This value for MHD_SAT is more in line with the expected results for late type spiral galaxies with solar metallicity of $2 \times 10^9$ yr (Bigiel et al. 2008).

The increase in cold gas and $\tau_{\rm dep}$ in the MHD models leads to the similarities between the SFRs. This is discussed in detail in section 4.3.

## 4 DISCUSSION

### 4.1 Magnetic Field

We have run three separate models to look at the effects a B-field will have on star formation in low metallicity environments. Two models started with a near primordial B-field strength of $10^{-6}\mu$G so the field was amplified by the galactic large and small scale dynamo, and the other with a higher field B-field of $10^{-2}\mu$G. The primordial models reach a saturation of $\sim 5\mu$G across the disc and the higher field model, MHD_SAT reaches saturation of $\sim 10\mu$G averaged across the star-forming disc with the toroidal component being dominant in all models. We also see that the strength of the B-field is related to the density, with the densest regions showing the highest strengths. We find a relationship of the form $|B| \propto n^{\alpha}$ with $\alpha \sim 0.5$, somewhat shallower than the value of $\alpha \sim 2/3$ found by Crutcher et al. (2010) at high densities. We leave a more detailed analysis of this result to future work.

During the growth of the B-field in MHD_10 there are periods where the field strength doesn't increase. Comparing these times to the star formation in Figure 7 they occur at the same time we see a drop in the SFR for this model shown by the blue line. The field only begins to grow again when the SFR rises. Star formation is associated with local collapse and subsequent feedback, which can twist and bend field lines, and energetic feedback that drives strong vorticity in the hot gas (Balsara et al. 2004; Gent et al. 2021, 2022), transferring energy into the field. Consequently, since there is variable star formation there is also variable field growth.

---

[1] We include fully molecular sinks in Appendix B

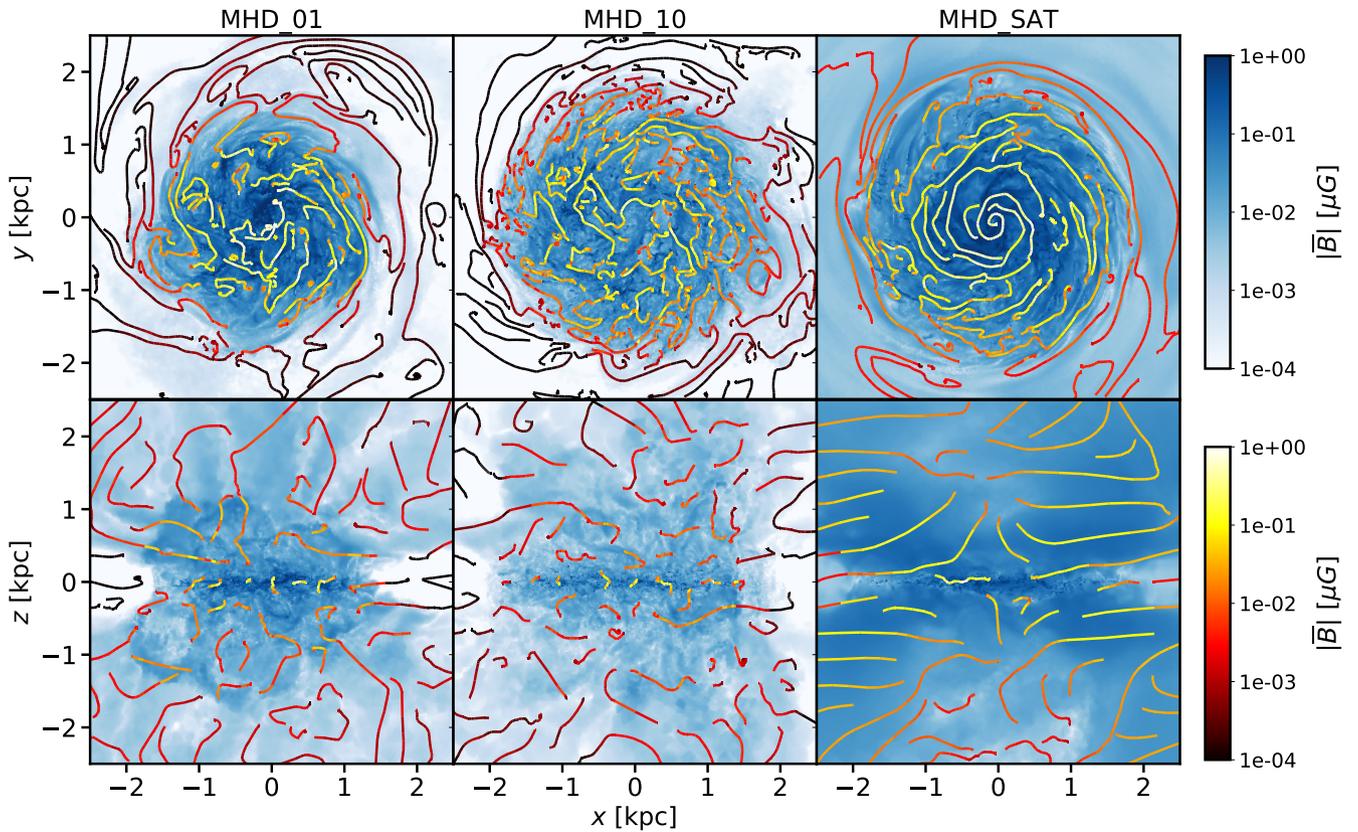

**Figure 5.** The absolute magnitude B-field strength (worked out from the summed squares of the 3-dimensional field vectors) and morphology at 1000 Myr. From left to right we show models MHD_01, MHD_10 and MHD_SAT. The top plots are face-on to the disc and the bottom plots are edge-on. The stream lines represent the field lines above $10^{-4}$ $\mu$G and are coloured to represent the strength of the field. Model MHD_SAT shows a much more ordered B-field compared to the models where the field was allowed to grow.

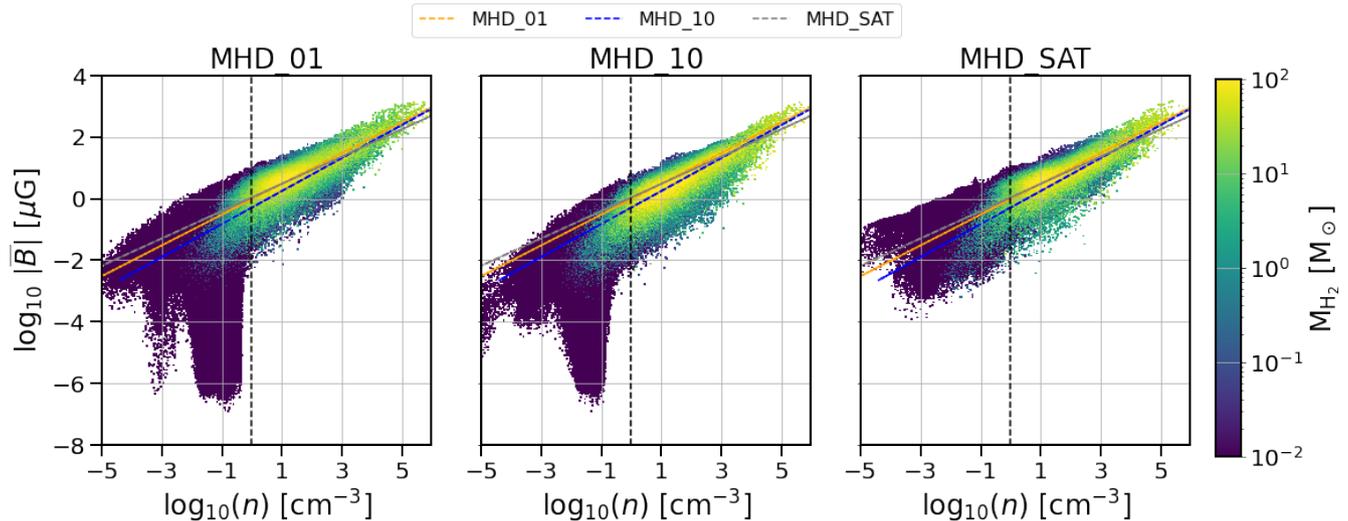

**Figure 6.** Mass-weighted absolute B-field ($|B|$) and number density phase plots at 1000 Myr. The dashed lines show the slope for each model for the cells with $|B| \geq 10^{-3}$ $\mu$G. All three models show the same upward trend in $|B|$ with number density. MHD_01 and MHD_10 both show very weak fields at low densities, below $n = 1$ cm$^{-3}$, corresponding to the regions left of the black dashed line. The cells with very weak fields are typically found far outside of the star-forming disk, where little dynamo action has occurred and the field strengths largely reflect their initial values. When analyzing the growth of the mass-weighted $|B|$, as in Figure 4, we exclude this region. This behaviour is not seen in MHD_SAT, as here even cells far outside the disk begin with a high $|B|$.

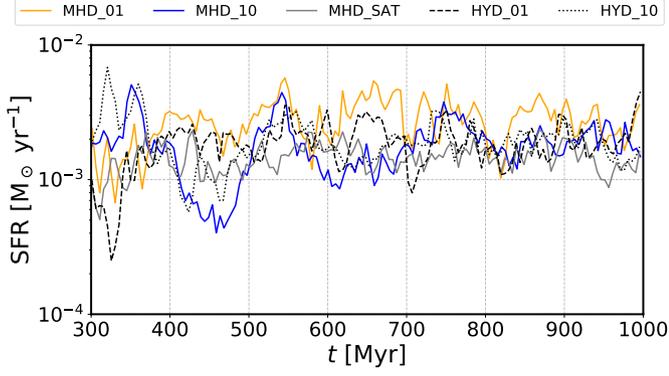

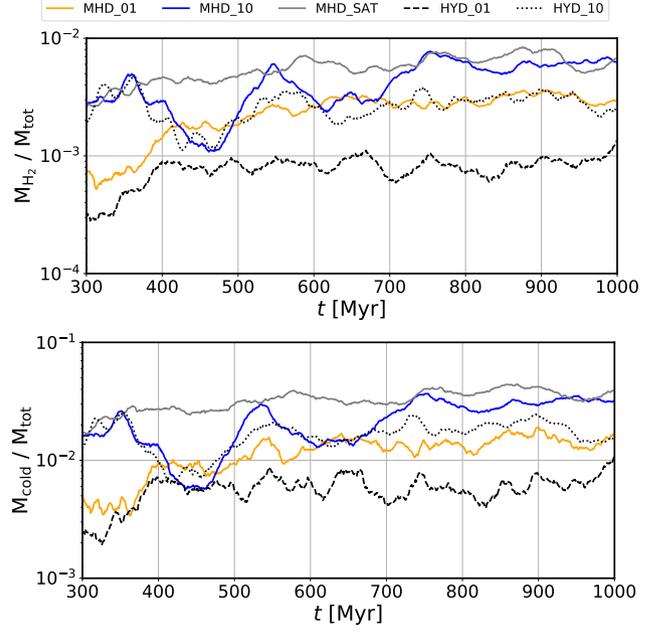

**Figure 7.** The star formation rate for all models shown over the steady state period. It is clear that the SFR is not suppressed by the addition of a magnetic field. There is no noticeable change across the models. In the MHD_SAT model we do see less stochasticity in the SFR, the bursty nature is reduced.

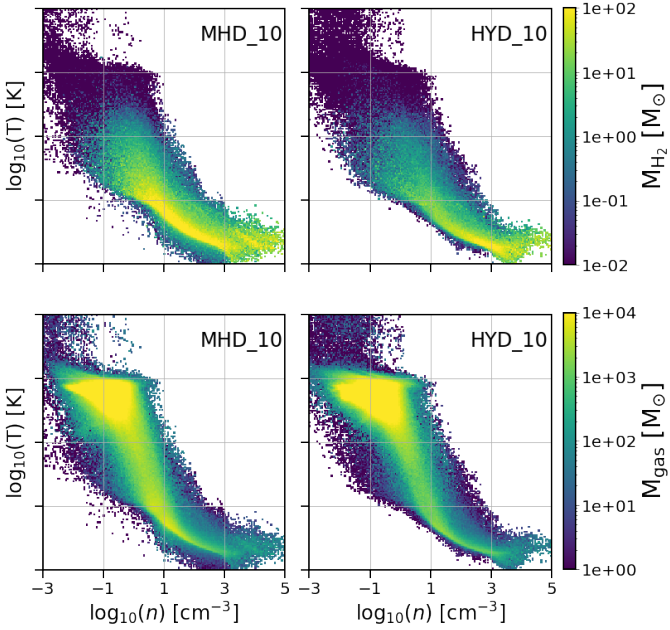

**Figure 8.** $H_2$ mass-weighted phase diagrams at $t = 1$ Gyr for MHD_10 and HYD_10. The majority of the $H_2$ mass lies in the cold (T < 100K) gas. In MHD_10 there is a much larger mass of $H_2$ in the cold phase compared to HYD_10.

**Figure 9.** Mass fraction of $H_2$ and cold gas (T < 100K) (MHD_01 = blue, MHD_10 = red, MHD_SAT = green, HYD_01 = grey and HYD_10 = black). All 3 MHD models show a higher mass fraction in both the $H_2$ and cold gas compared to the hydrodynamical models. The MHD_SAT model shows a more steady mass fraction, whereas MHD_01 and MHD_10 show some stochasticity and have a slight upward trend.

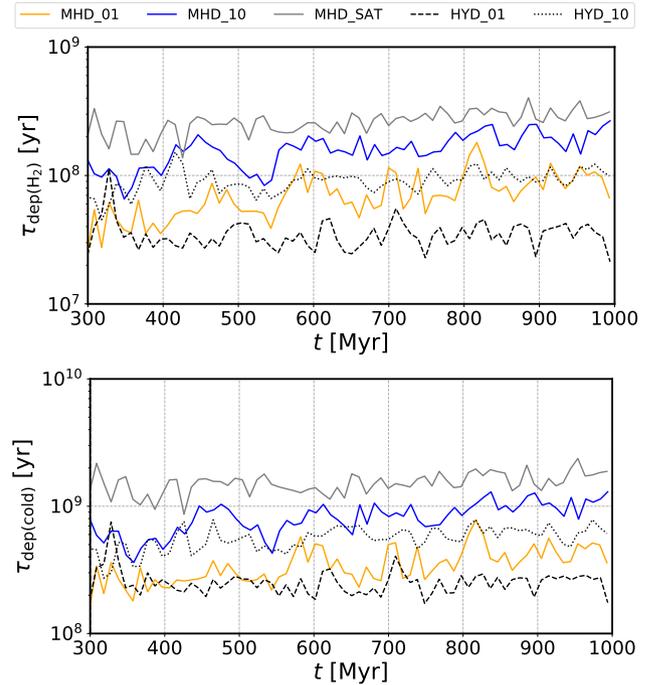

**Figure 10.** Depletion time ($\tau_{dep}$) for $H_2$ and cold gas as a function of time for all models (MHD_01 = blue, MHD_10 = red, MHD_SAT = green, HYD_01 = grey and HYD_10 = black). Both the $\tau_{dep(H_2)}$ and $\tau_{dep(cold)}$ increase over time in MHD_01 and MHD_10. MHD_SAT is steady and higher than all models over the entire steady state.

As the SFR is not suppressed by the field, the question is if the B-field is dynamically important? To consider this we look at the plasma-$\beta$, the ratio of thermal pressure to magnetic pressure, in the star forming disc:

$$\beta = \frac{P_{th}}{P_m},    (5)$$

where $P_{th}$ is the gas pressure, $P_{th} = (\gamma - 1)\rho_0 \times u_{th}$, where $u_{th}$ is the specific thermal energy of the cell, $\rho_0$ its mass density and $P_m$ is the magnetic pressure, $P_m = B^2/8\pi$. When $\beta$ is greater than 1 the thermal pressure of the gas is dominant, and below 1 the magnetic pressure is dominant.

Looking at molecular mass-weighted plasma-$\beta$ in relation to number density at 1 Gyr, Figure 11, the discs in MHD_01 and MHD_10

are dominated by thermal pressure compared to magnetic pressure, especially in the diffuse gas. For model MHD_SAT thermal pressure is still dominant but to a much lesser degree. However in the densest regions the B-field becomes more important than thermal pressure and will act as an additional supportive force against gravitational collapse across field lines in the cold, dense clouds. In the low-$\beta$ regions gravitational collapse is constrained to follow the field lines. Star formation can only proceed when the mass locally exceeds the critical mass-to-flux ratio (e.g. Ibáñez-Mejía et al. 2022).

In Figure 12 we show the radial profiles at 1 Gyr for both the magnetic energy density, $E_m = |B|^2/8\pi$, and kinetic energy density $E_k = 1/2\,\rho\sigma_{v_{turb}}^2$. $E_k$ is calculated using a method similar to Beck (2015) where the mass-weighted velocity dispersion of the turbulent gas ($\sigma_{v_{turb}}$) is calculated from the cylindrical velocity vector components, $v_r$, $v_\theta$ and $v_z$, which as the circular component is roughly constant removes the systematic component from rotation. This is calculated for each cell in an annuli of 5 pc from 0 to 2.5 kpc, creating 500 bins. Each bin is then mass-weighted to give the mass-weighted $\sigma_{v_{turb}}$. The density ($\rho$) and mass-weighted absolute B-field ($|B|$) are binned in the same manner. Each radial bin is then plotted, giving the radial profile seen in Figure 12.

The $E_m$ has not saturated the disc and the $E_k$ is dominant. This shows that the B-field is dynamically less important overall compared to the turbulent flow within the star-forming disc at this time. We also note that the energy densities follow roughly the same decreasing trend radially outwards. This is in opposition to what is seen in larger galaxies with more dominant fields (Beck 2015), where $E_m$ decreases more slowly with radius than $E_k$.

Figure 13 compares $E_m$ (right column) and $E_k$ (middle column) to the number density in the star-forming disc at 1 Gyr for each model. To see if the models reached equipartition we plot $E_m/E_k$ (left column). This is plotted in log units, so a value of 0 corresponds to equipartition. The black dashed line shows where we make the density cut in Figure 4. In MHD_01 and MHD_01, there is a drop in $E_m$ at low densities which we do not see in MHD_SAT. This is due to MHD_SAT starting with a higher field strength across the disc. Once $E_m$ reaches 10% of $E_k$, the B-field is expected to saturate (Pakmor et al. 2017). We find that the MHD_SAT model is the only model close to this at ~ 9.38%, with MHD_01 at ~ 0.40% and MHD_01 at ~ 0.23%. This is likely to arise from the contribution of the mean ratio made by diffuse gas at the edge of the star-forming disc where the B-field is very weak in MHD_01 and MHD_10.

### 4.2 Molecular Gas Fraction

Where the morphological differences are most substantial between the hydrodynamical models and MHD models is in the distribution and amount of $\Sigma_{H_2}$. Looking at the difference between MHD_01 and HYD_01 in the top two rows of Figures 1 and 2 it is clear that there is more $H_2$ present in MHD_01, both across the disc and when viewed edge-on. Though not as clear, the same is true for MHD_10 and MHD_SAT compared to HYD_10. In these models there are larger and more clumps of $H_2$ spread across the disc, and a small broadening of the vertical distribution.

To investigate the origin of the extra $H_2$ in the disc we first test if the total amount of gas has increased. As we are using ideal MHD the gas is tied to the B-field, and so as the large scale dynamo builds up ,the B-field may stop gas from escaping the disc. Within the star forming disc ($r = 2.5$ kpc, $|z| < 0.4$ kpc) at year 0 all models have a mass of $5.04 \times 10^7\,M_\odot$.

Mass is lost in models MHD_10, MHD_SAT and HYD_10, as shown in Figure 14. The loss in mass before the steady state arises from the initial burst in star formation and the subsequent SNe going off in quiescent gas causing gas to be ejected from the disc. The mass loss appears to be controlled primarily by the variation in metallicity, UV fields, and cosmic ray ionization rate between the models rather than by the magnetic fields. Most of this gas mass is lost through galactic outflow and the initial burst of star formation, and not consumed by star formation during the steady state. As the gas mass is broadly similar in both HD and MHD cases with similar parameters, it can not be the origin of the discrepancy in the molecular gas fraction.

At the beginning of the steady state period the mass fractions of molecular hydrogen and cold gas in the MHD models is similar to that of the HD models with similar parameters (Fig. 9). It is only when the B-field approaches saturation that we see them start to diverge. In MHD_SAT this happens quickly, within 100 Myr of the start of the steady state, and becomes relatively stable, whilst MHD_01 is stochastic up to 700 Myr, reaching the same mass fraction as MHD_SAT. The mass fractions of MHD_01 also grow over the course of the steady state, reaching stability at ~ 500 Myr. As the B-field grows so does the molecular mass fraction suggesting a connection between two.

But where does this extra molecular gas lie? We have plotted the cumulative mass of $H_2$ ($M_\odot$) against number density ($cm^{-2}$) for each model in Figure 15. We can see the both MHD_SAT and MHD_01 have more $H_2$ across all number densities in all plots compared to the comparable HD models. MHD_10 shows an increase in mass from at least 750 Myr.

This is in agreement with the work of Pardi et al. (2017) who show that a weak ($6 \times 10^{-3}\mu G$) magnetic field produces a greater mass fraction of molecular and cold gas compared to a non-magnetic model in a model of a giant molecular cloud with solar metallicity and UV-field strength. In contrast they found that a strong field supports the molecular cloud against gravitational collapse, which reduces the molecular and cold gas mass fractions as expected. They conclude this is due to systematic effects of their SNe positioning and magnetic tension in clumps preventing the dense gas from being dispersed.

The analysis of the plasma-$\beta$ in Figure 11 shows that the magnetic field becomes as important as thermal pressure at number densities of above 100 $cm^{-3}$ where the gas may become molecular.

In the weak-field scenario, molecular clouds are somewhat controlled by turbulent flows in the gas (Mac Low & Klessen 2004). To see if there is sufficient magnetic pressure support to oppose gravity in the dense gas we now consider the mass-to-flux ratio ($M/\phi$) in Figures 16 and 17. To calculate this for the projections in Figure 16 we use the method from Crutcher (2004)

$$M/\phi = 7.6 \times 10^{-21} N(H_2)\,/\,|B|, \tag{6}$$

where $N(H_2)$ is the column density of $H_2$ in $cm^{-2}$ and $|B|$ is the total field strength in microgauss from the line of sight projection in Figure 5. We make a cut at $N(H_2) = 10^{15}\,cm^{-2}$ to trace dense molecular gas. If we were to take this plot at face value then we would naively think that only the densest molecular gas has a supercritical mass-to-flux ration, i.e. $M/\phi > 1$. However, if we directly calculate $M/\phi$ for the molecular gas

$$M/\phi = \frac{M_i}{\pi r_i^2 |B_i|}, \tag{7}$$

where $M_i$ is the mass of molecular gas within a cell $i$ in grams, $r_i$ is the radius of cell $i$ in centimetres and $|B_i|$ is the absolute strength of the B-field in Gauss, shown in Figure 17 we see that a large fraction of the molecular gas is above $M/\phi = 1$, shown by the black line on the plots. This means that it is supercritical and dominated by gravity so will collapse and form stars. The more diffuse, warmer gas is dominated by the B-field. This stops the gas from collapsing and keeps it in the clouds allowing it to accumulate, hence the larger gas fraction we see.

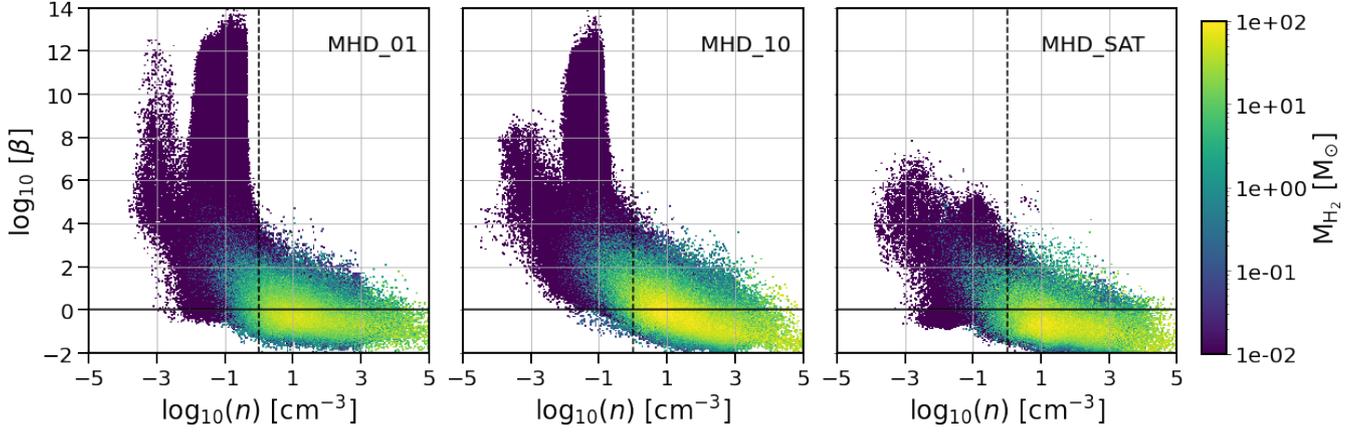

**Figure 11.** Plasma $\beta$ versus number density for each of the MHD models at 1 Gyr. From left to right, we show MHD_01, MHD_10, MHD_SAT. The solid line shows a plasma-$\beta$ of 1. Below this line, the magnetic pressure dominates over the thermal pressure. The black dashed line indicates where we make the cut when analyzing the mass-weighted $|B|$. In all three models, in the densest regions the B-field has become dominant. In these regions the clouds will be supported from collapse across the field, while in the diffuse regions the thermal pressure is dominant.

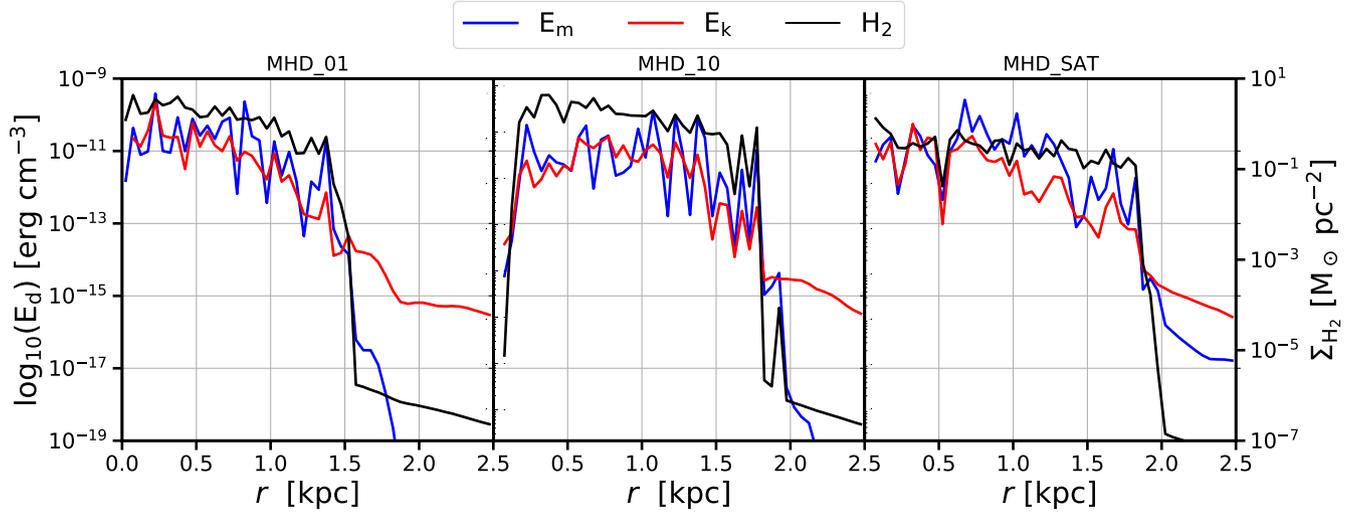

**Figure 12.** Magnetic ($E_m$, blue line) and kinetic ($E_k$, red line) energy densities and the surface density of $H_2$ ($\Sigma_{H_2}$, black line) shown as a function of galactocentric radius for each MHD model, left - MHD_01, middle - MHD_10, right - MHD_SAT taken at 1 Gyr. We can see that $E_m$ has yet to saturate and $E_k$ is still dominant in all 3 models. We can also see that the energy densities trace the $\Sigma_{H_2}$ well, dropping off at the edge of the star-forming disc.

Figure 17 gives a more complete picture of $M/\phi$ as it shows the 3D distribution of the gas and the magnetic field, whereas the projection in Figure 16 does not take into account the bulk mass of the molecular gas and assumes a relationship between surface number density and mass in line with how observations would calculate this. In the cell-by-cell plot we can more accurately calculate the molecular mass.

Though the molecular clouds are not supported against collapse the magnetic pressure will slow it. This should lower the SFR, but with the increased mass fractions in both cold and molecular gas the SFR actually increases cancelling out the expected suppression.

### 4.3 Star Formation

Previous studies (Price & Bate 2009; Federrath & Klessen 2013; Hennebelle & Inutsuka 2019; Krumholz & Federrath 2019) have shown that B-fields suppress star formation in individual molecular clouds as the field slows the collapse of the gas. However, given the increase in $\Sigma_{H_2}$ that we find in our magnetized models, if the molecular gas was causally linked to star formation (see Paper I for a full discussion of this) then one would expect to see an increase in SFR. However, as will be discussed here, this is not the case, we see neither a large scale increase or suppression.

The star formation rate in all our dwarf galaxy models is $\sim 2 \times 10^{-3}$ $M_\odot$ $yr^{-1}$. There is some minor variation from this (Fig. 7), with MHD_01 having an increased rate compared to HYD_01 and MHD_10 and MHD_SAT having a slightly decreased rate compared to HYD_10, but overall the star formation rates are not suppressed as would be expected from studies of magnetic fields in molecular clouds.

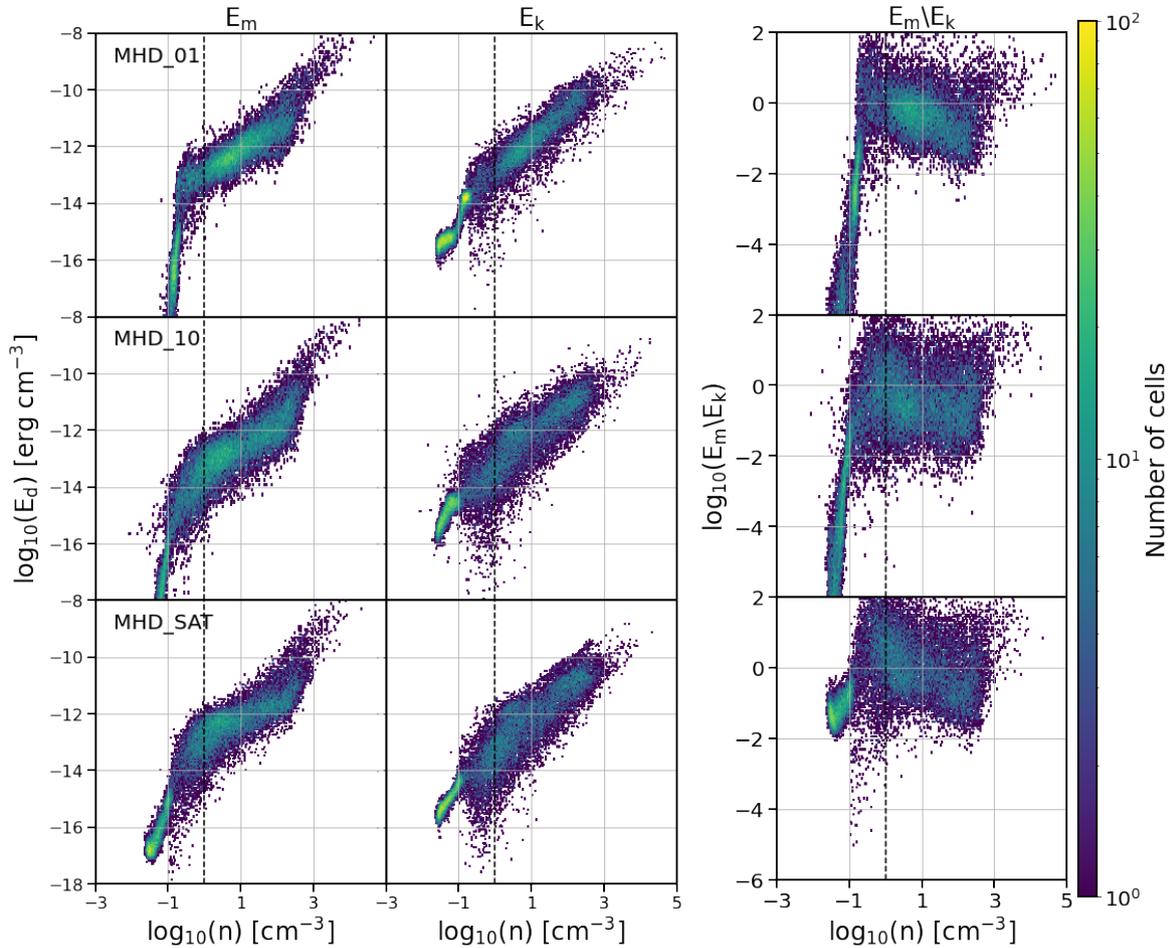

**Figure 13.** Distributions of magnetic energy density ($E_m$) (*left*) and kinetic energy density ($E_k$)(*center*) as a function of the number density $n$ at 1 Gyr for models MHD_01 (*top row*), MHD_10 (*middle row*), and MHD_SAT (*bottom row*). On the *right* we show the ratio of magnetic to kinetic energy (note that this is the inverse of a kinetic plasma $\beta$). The black dashed line in each plot is the $n = 1$ cm$^{-3}$ cut we make when analyzing the mass-weighted B-field strength. Although the plasma $\beta$ is low in dense gas, this figure shows that the kinetic energy still dominates in those regions, probably from hierarchical gravitational collapse (e.g. Ibáñez-Mejía et al. 2022).

On the other hand, as discussed in section 4.2 there is more H$_2$ and cold dense gas in the MHD models, and as shown in Whitworth et al. (2022) it is the cold dense gas that determines the star formation. However, as shown in Figure 9, the cold gas fraction is also higher in the magnetised models. From this reasoning one would actually expect the star formation to be higher in the MHD case.

One way to resolve this contradiction is to consider the principle of self-regulation. As summarised by Ostriker & Kim (2022a) the mid-plane pressure is in vertical dynamical equilibrium with the weight of the ISM. This is because the main source of energy supporting the disc comes from supernova explosions and stellar feedback, meaning that the forces must form an equilibrium where the local star formation should correlate with the mid-plane pressure (Kim et al. 2013). As the mass of the galactic disc is similar in the hydrodynamic and MHD runs, the star formation rate is also similar. The cold gas fraction and molecular mass fraction is therefore higher because the field has made the process of star formation less efficient. In principle, adding a magnetic field should add an additional form of pressure support (magnetic pressure), which implies a priori that less star formation is needed to support the disk. However, as shown in Figure 13 the magnetic energy density is smaller (or at best equal) to the kinetic energy density on these scales, so the kinetic energy dominates the vertical support in both the hydro and MHD runs. Therefore, adding the field in practice has little impact on the vertical support and hence little impact on the SFR. This hypothesis is supported by the increased depletion times of the MHD models compared to the HD models shown in Figure 10.

In effect we see a balancing between the slowing of the collapse of the gas in the ISM and therefore a slowing of local star formation and an increase in the cold and molecular star forming gas mass fraction (which will increase the global star formation). *Globally* there is consequently little change in the star formation rate between the hydrodynamical models and MHD models.

### 4.4 Velocity dispersion

Given that the star formation is similar in both cases we investigate the gas velocity dispersion to probe the effect of the magnetic field on SN feedback, as this is the main source of turbulence in our simulations. For this we look at four different epochs, 250 Myr, 500 Myr, 750 Myr and 1 Gyr. Two components of the velocity dispersion are considered, perpendicular to the disc $\sigma_{v_p}$ and radial $\sigma_{v_r}$. We

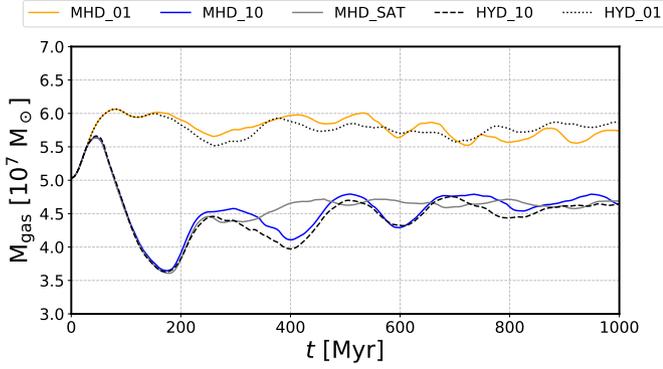

**Figure 14.** Total gas mass ($M_\odot$) in the star forming disc over the full 1 Gyr of the simulation. The three models with $Z_\odot = 0.10$ (MHD_10, MHD_SAT and HYD_10), before the steady state begins at 300 Myr we see an increase in mass and then a decrease which is caused by the initial burst in star formation and the subsequent supernovae going off in quiescent gas.

calculate the mass-weighted velocities

$$v_r = \frac{(v_x \cdot x + v_y \cdot y)}{r} \quad (8)$$

and

$$v_p = |v_z|, \quad (9)$$

where **x**, **y** and **z** are the vector positions of the velocity vectors $v_x$, $v_y$ and $v_z$ and $r = (x^2 + y^2)^{1/2}$ is the cylindrical radius. $v_p$ is the perpendicular velocity taken as the vector in the $z$-plane moving towards or away from the plane of the disc.

In Figure 18 we plot the mass-weighted velocity dispersion, $\sigma_{v_p}$ (top) and the radial velocity, $\sigma_{v_r}$ (bottom), as radial profiles out to 2.5 kpc. From these plots we can see that the $\sigma_{v_r}$ for all models is similar with no large scale deviations, even at 250 Myr, before the steady-state has begun.

Looking at $\sigma_{v_p}$ the gas in the MHD models has less vertical dispersion than in the HD models. At all times the hydrodynamical models show a larger $\sigma_{v_p}$. This larger dispersion is produced by the SNe feedback that is dominating the gas in HYD_01 and HYD_10. In the MHD models the gas accelerated by SNe has to work against the magnetic field lines and so the perpendicular motion of the gas is reduced and the gas leaves the disc with a lower velocity. The SNe rate in the models are all of the same order $\sim 10^1$ Myr$^{-1}$.

Over time the $\sigma_{v_p}$ of the hydrodynamical models does reduce, but even at 1 Gyr HYD_01 is still a factor of 2 larger than MHD_01 and HYD_10 is a factor of $\sim 4$ larger than MHD_10 and MHD_SAT.

### 4.5 Small-scale or large-scale dynamo

It is well known that SSDs occur in the ISM (Gent et al. 2021), and are possibly active in dwarf galaxy evolution (Rieder & Teyssier 2016, 2017). The SSD should drive B-field growth at early times, but normally the LSD is expected to be the dominant amplification effect at late times. It is expected that fast exponential amplification on the eddy turnover timescale is driven by the turbulent SSD, while the LSD only drives exponential growth on the galactic rotation timescale.

In our models, for the LSD to become dominant the disc would need to have completed numerous rotations. The discs in our simulated dwarf galaxies have an average rotational velocity of around 35 km s$^{-1}$, varying slightly between models. The orbital periods that arise from this are $\sim 400$ Myr. These long orbital periods do not allow for many rotations, even in our 1 Gyr simulations, for the LSD to grow the field.

If the LSD were to be active the energy needed to amplify the field and maintain it would come from the kinetic energy of the disc rotation. Though there is a small increase in orbital periods between the HD and the MHD models, a 4 Myr increase in the 10% models and a 12 Myr increase in the 1% models, this is not a large discrepancy. This means that the SSD is most likely still the most significant form of amplification in all the MHD models at the end of the simulations.

The amplification from the SSD seen in our models is slower than that seen in other works studying small scale field growth at higher uniform numerical resolution (Pardi et al. 2017; Rieder & Teyssier 2017; Gent et al. 2021) where the amplification takes place on the order of a few tens of megayears. However, these models resolve the hot gas where the field grows fastest with parsec to sub-parsec resolution, while our model only resolves that gas with 10 pc or worse resolution. At larger, Milky Way scales, even more poorly resolved models do show that the B-field grows exponentially, but over a longer period, from one to a few gigayears (Pakmor et al. 2014, 2017; Steinwandel et al. 2020).

We conclude that the field amplification we observe is predominantly driven by the turbulent SSD. We would only expect the LSD to act significantly over many rotation periods.

### 4.6 Caveats

As in Paper I, we lack photoionisation feedback from individual HII regions. We also do not track the contribution of individual star-forming regions to the softer UV responsible for photodissociating $H_2$ and heating the ISM via the photoelectric effect. Instead, we use a fixed FUV field and account for local attenuation of this field by the gas. By excluding photoionisation we are likely over-estimating the amount of atomic hydrogen present in our simulated galaxies and under-estimating the ionized gas fraction. This could affect our results in two ways: the higher pressure of the ionized gas compared to the warm neutral gas implies additional pressure support and hence potentially slower collapse; and we are missing the momentum input associated with photoionisation, which can be considerable (Ostriker & Kim 2022b), although Jeffreson et al. (2021) see no effect on SF in Milky Way like models when they include photoionisastion feedback in comparison to only SNe feedback.

Another point to bear in mind is that in our treatment of the magnetic field we use the Powell cleaning method to control the field divergence, rather than using a method that conserves $\nabla \cdot \mathbf{B}$ by construction. This compromise is forced on us by computational efficiency concerns, but we note that the Powell method has been proven in many studies to be accurate for modelling galactic scale magnetic fields (Pakmor & Springel 2013; Pakmor et al. 2014; Van de Voort et al. 2021).

We also note that the field our model includes only numerical resistivity. Non-ideal MHD processes have been shown to be dominant in the forming of stellar cores and protostellar discs (Wurster & Lewis 2020a,b), but as our resolution is unable to resolve these details, using an ideal MHD model should be appropriate.

The original orientation of the field is in the $z$-plane, perpendicular to the disc. As discussed in section 3.2, the field morphology quickly becomes toroidally dominated. Starting with a toroidal field would reduce the time for the field to become aligned with the rotation of the disc, but beyond that, we feel that there would be little impact as the toroidal component quickly becomes dominant.

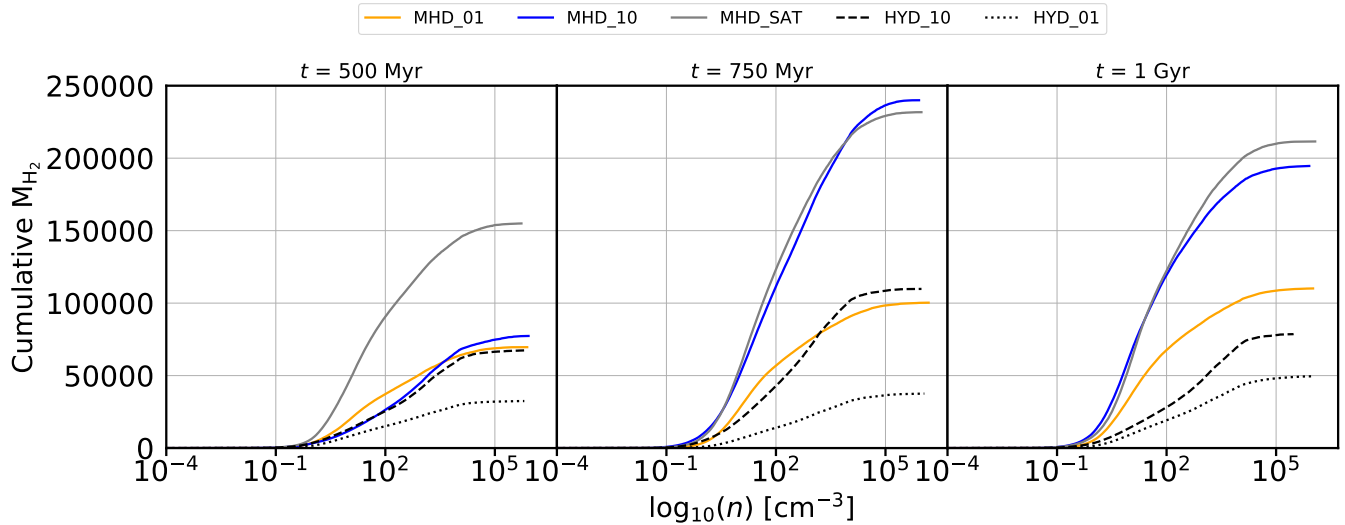

**Figure 15.** The cumulative mass of $H_2$ ($M_\odot$) in relation to number density (cm$^{-2}$) for each model at 500 Myr (left panel), 750 Myr (middle panel) and 1 Gyr (left panel). We can see that at all time the MHD models across all densities have a greater mass of $H_2$ than the comparable HD model. At later times the seperation is more pronounced.

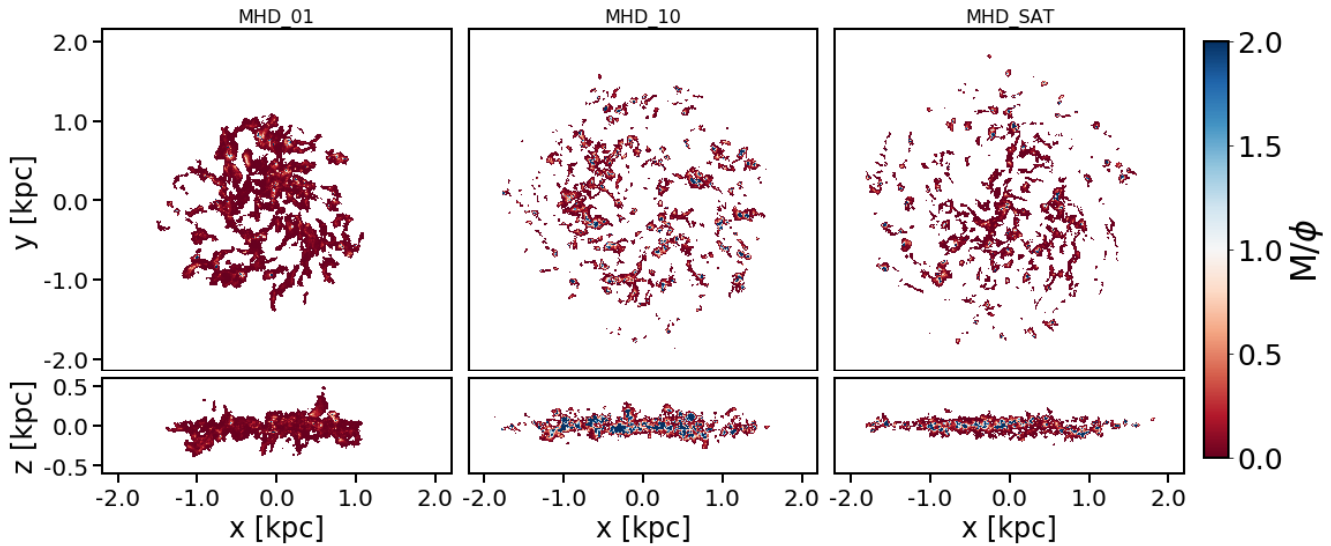

**Figure 16.** The mass-to-flux ($M/\phi$) ratio for each model at 1 Gyr. The left panel is MHD_01, middle panel is MHD_10 and right panel shows MHD_SAT. We have make a cut to show only the $M/\phi$ where $N(H_2)$ is greater than $10^{15}$ cm$^{-3}$ so we can see the effects of the B-field and gravitational forces on the densest molecular gas. These plots show that only in the densest regions is the $M/\phi$ greater than 1. This shows the weak-field model is active.

## 5 CONCLUSIONS

We modelled five isolated dwarf galaxies, two HD and three MHD, using the moving mesh code AREPO. The HD models were taken from Paper 1, whilst the MHD models are new. The initial field seed was set to a primordial field strength of $10^{-6} \mu G$ for two models and allowed to evolve, and one model started with a close to saturation seed field of $10^{-2} \mu G$. We varied the metallicity, dust-to-gas ratio, UV field strength and cosmic-ray ionisation rate between 1% and 10% of solar values. All models were run for 1 Gyr and we used a

defined steady state from paper 1 of 300 Myr to 1 Gyr. The main results are summarised below:

- Unexpectedly, the global star formation rate in the MHD models is not suppressed compared to the HD models. This is in opposition both to metal-rich cloud scale models that start with an evolved and saturated B-field, and cosmological models of larger, more metal rich galaxies.

- Morphologically, the MHD models have a broader distribution of $H_2$ both radially and vertically, in agreement with current literature.

- All models are dominated by warm $H_I$ in the disc. However there is a noticeable increase in the amount of $H_2$ and cold (T< 100K) gas in the MHD models. The cold gas fraction rises from 1.66% in model HYD_10 to 3.24% in saturated MHD model MHD_SAT, whilst the molecular hydrogen fraction increases from 0.27% to 0.56% between the same models.

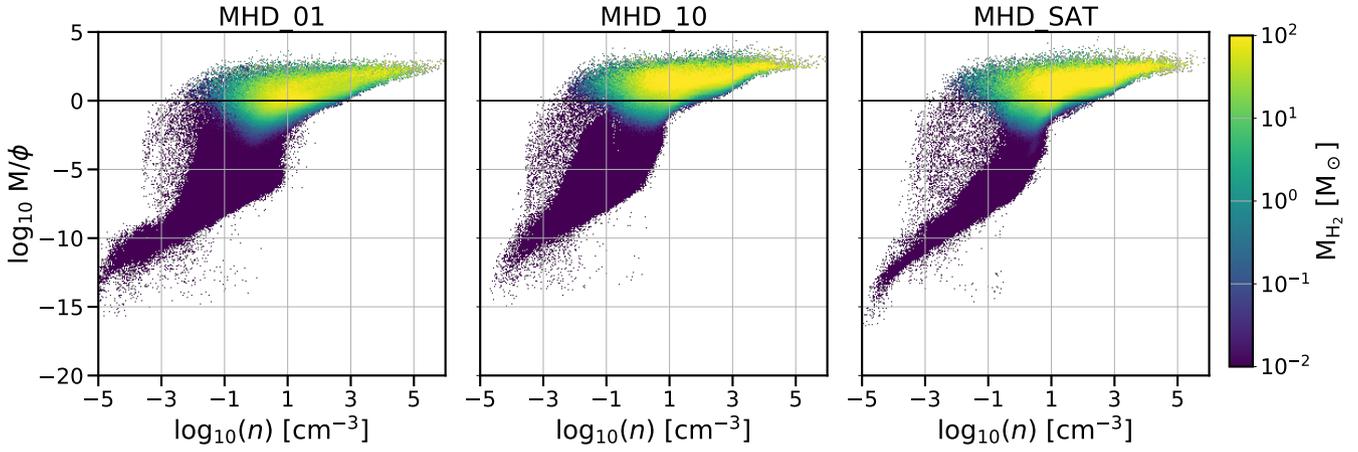

**Figure 17.** The mass-to-flux ($M/\phi$) ratio for each model at 1 Gyr against number density weighted by $H_2$ mass. The left panel is MHD_01, middle panel is MHD_10 and right panel shows MHD_SAT. In the dense regions, above n = $100\,\mathrm{cm}^3$, the molecular gas is all supercritical with $M/\phi > 1$, shown by the black line in the plots. Only in the diffuse gas is there some subcritical $H_2$.

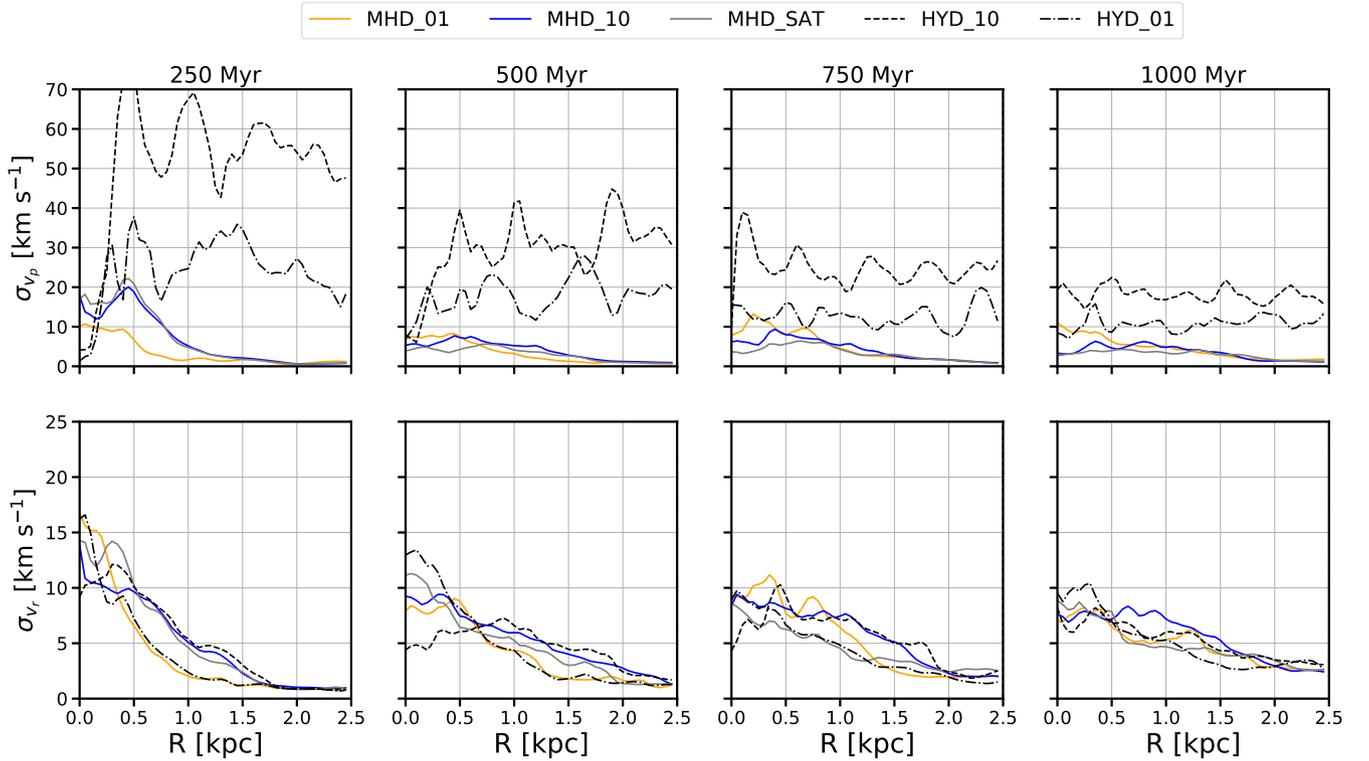

**Figure 18.** Radial profiles of the mass-weighted perpendicular velocity dispersion , $\sigma_{v_p}$, (top) and the mass weight radial velocity dispersion, $\sigma_{v_r}$, (bottom) at 250 Myr, 500 Myr, 750 Myr and 1 Gyr for all models (MHD_01 = blue, MHD_10 = red, MHD_SAT = green, HYD_01 = grey and HYD_10 = black). In the MHD case $\sigma_{v_p}$ is suppressed across the disc in the steady-state period.

- The depletion time for both $H_2$ and cold gas in the MHD models is longer than for the corresponding HD models. This occurs because the magnetic field locally supports individual molecular clouds against collapse into star-forming regions.

- The mass weighted B-field saturates across the disc at $\sim 5\mu$G for the MHD_01 and MHD_10 models and at $\sim 10\mu$G for the MHD_SAT model, but saturates at different times in each model.

- We find a power-law relationship between number density of the gas and B-field strength in all models, with a power-law exponent of $\sim 0.5$.

- The toroidal component of the B-field becomes dominant after $\sim 50$ Myr in all models.

- The magnetic energy density approaches equipartition to the kinetic energy within the disc. However, the ratio between the magnetic and kinetic energy density varies and is somewhat dependent on the density of the gas.

## ACKNOWLEDGEMENTS

We thank the members of the AREPO ISM group and EcoGal Large Scale Structures group for discussions and insightful comments on the coding and science goals in this paper. DJW also thank the postgraduate groups at Jodrell Bank Centre for Astrophysics for interesting and engaging science discussions and supportive comments; Volker Springel for access to AREPO; and we also thank the anonymous referee for comments that helped improve the paper. DJW is grateful for support through a STFC Doctoral Training Partnership. RJS gratefully acknowledges an STFC Ernest Rutherford fellowship (grant ST/N00485X/1). DJW, SCOG, RT, JDS and RSK acknowledge funding from the European Research Council via the ERC Synergy Grant "ECOGAL – Understanding our Galactic ecosystem: From the disc of the Milky Way to the formation sites of stars and planets" (project ID 855130). They also acknowledge support from the DFG via the Collaborative Research Center (SFB 881, Project-ID 138713538) "The Milky Way System" (sub-projects A1, B1, B2 and B8) and from the Heidelberg cluster of excellence (EXC 2181 - 390900948) "STRUCTURES: A unifying approach to emergent phenomena in the physical world, mathematics, and complex data", funded by the German Excellence Strategy. M-MML acknowledges support from US NSF grant AST18-15461, and hospitality from the Institut für Theoretische Astrophysik at U. Heidelberg.

This work used the DiRAC COSMA Durham facility managed by the Institute for Computational Cosmology on behalf of the STFC DiRAC HPC Facility (www.dirac.ac.uk). The equipment was funded by BEIS capital funding via STFC capital grants ST/P002293/1, ST/R002371/1 and ST/S002502/1, Durham University and STFC operations grant ST/R000832/1. DiRAC is part of the National e-Infrastructure. The research conducted in this paper used SciPy (Virtanen et al. 2020), NumPy (Van der Walt et al. 2011), and matplotlib, a Python library used to create publication quality plots (Hunter 2007).

## DATA AVAILABILITY

All data is available upon request to the authors of this work.

## APPENDIX A: RESOLUTION TEST

To check whether the results we found are a result of numerics, especially in regards to the speed of the growth of the field we ran model MHD_10 at a lower resolution, model JR4_10. For this test we ran with a set the Jeans Refinement criteria to be set at 4 cells instead of 8. We also increased the target cell mass from $50\,M_\odot$ to $500\,M_\odot$.

With respect to cell size, Figure A1 shows the cell radii for JR4_10 in the histogram plot, the contours show the data from MHD_10 as a comparison. We can see that the cell radii is smaller in JR4_10, with fewer cells reaching the sink creation density threshold. This has an effect on the morphology, Figure A2, where we note a less disrupted HI surface density ($\Sigma_{HI}$) but a similar filling factor of dense $H_2$ in the $H_2$ surface density ($\Sigma_{H_2}$) of 3.3%, though this is less disrupted like the $\Sigma_{HI}$.

Comparing the growth of the B-field between the two models, Figure A3 we note that the field in JR4_10 takes longer to reach saturation, $> 1$ Gyr. The B-field grows steadily after the initial burst reaching a strength of $10^{-1}\,\mu G$ at 1 Gyr int the total field strength.

Figure A4 compares the star formation rate between JR4_10 and MHD_10 over the steady state. On average the steady state star formation rate is the same between the models, with JR4_10 at $1.21\times 10^{-3}\,M_\odot yr^{-1}$ and MHD_10 at $1.84 \times 10^{-3}\,M_\odot yr^{-1}$. The SFR in JR4_10 is much more stochastic over time though.

We do see a slight decrease in the depletion time of $H_2$ ($\tau_{H_2}$) in JR4_10 to 58.1 Myr from 61.6 Myrs in MHD_10, Figure A5. This arises because we no longer have the resolution to resolve molecular gas at high densities. This leads to a smaller fraction of $H_2$, 0.10% in JR4_10 compared to 0.43% in MHD_10 over the steady state. Over the steady state the $\tau_{H_2}$ for JR4_10 is $7.91 \times 10^7$ yr and MHD_10 is $16.3 \times 10^7$ yr. There is no significant change in the cold gas depletion time ($\tau_{cold}$) between the models.

We conclude that the resolution has an effect on the growth of the

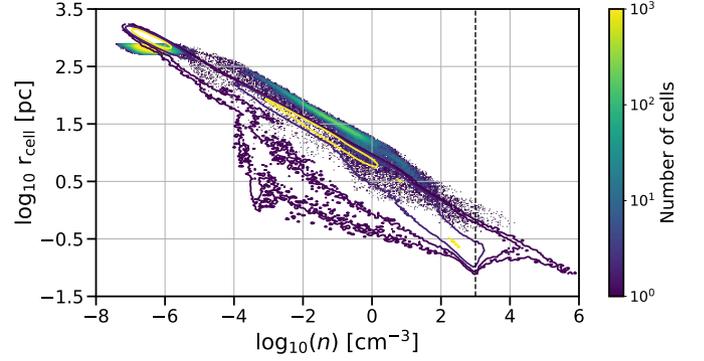

**Figure A1.** Cell radius as a function of number density. $r_{cell}$ is the radius of a sphere that has the same volume as the cell, taken at 500 Myr from model JR4_10. The contours trace the cell radius for MHD_10. We reach sub-parsec resolution at densities that are comparable to the sink formation threshold of $10^3$, $cm^{-3}$, shown by the dashed line, but we see little gas beyond this, with most of the cells sitting at large radii and low surface density.

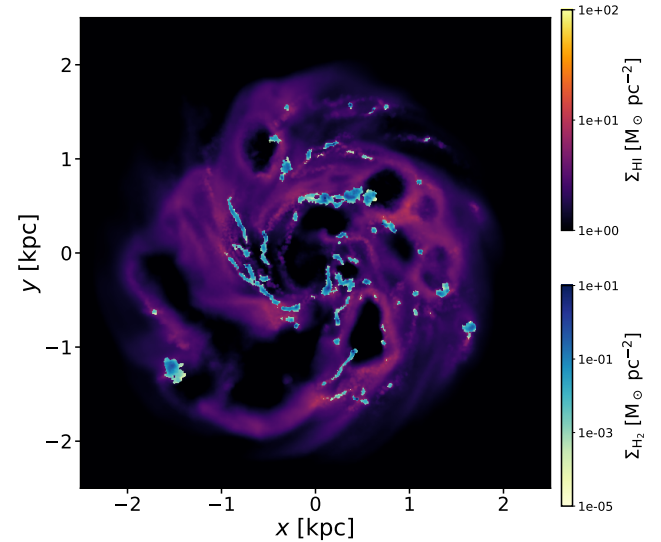

**Figure A2.** The HI surface density ($\Sigma_{HI}$) and $H_2$ surface density ($\Sigma_{H_2}$) for JR4_10 at 1 Gyr. When compared to model MHD_10 we can see the $\Sigma_{HI}$ is less disrupted and fewer $H_2$ clouds.

B-field, but only a minimal one. The same goes for the SFR, where we only see a small decrease in rate when we reduce the resolution. The larger effect is on formation of molecular gas, where we see less. The decrease in molecular gas has little impact on SFR as in low metallicity systems like these star formation is more closely tied to cold dense gas (Whitworth et al. 2022).

## APPENDIX B: MOLECULAR SINKS

In the bulk of this work we do not consider the gas trapped within the sinks, which amount's to 90% of the mass of the sink particle. We do this as we have no information on the chemical makeup of this gas. Here we show the $H_2$ mass fraction plot and $\tau_{dep(H_2)}$ for the MHD models if we were to consider the gas in sinks to be fully molecular.

When looking at the $H_2$ mass fraction, Figure B1, for the models over the steady state period we note that there is an increase when we include the gas mass of the sinks. The same is true for the depletion times, Figure B2. MHD_01 shows the largest difference, with an increase of a factor of $\sim 3$ compared to not including the sinks, Table B1.

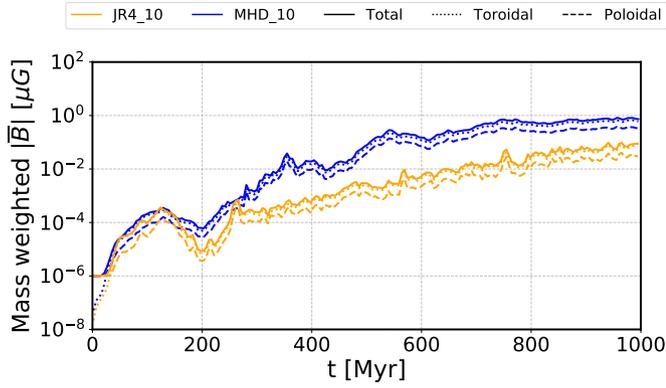

**Figure A3.** Growth of the mass-weight B-field ($|\overline{B}|$) for models JR4_10 (orange lines) and MHD_10 (blue lines). We plot the total (solid line), toroidal (dotted line) and the poloidal (dashed line) compents for each model. JR4_10 grows slower then MHD_10 and is more stochastic. At 1 Gyr they are of similar strength, $\sim 10^0 \mu$G.

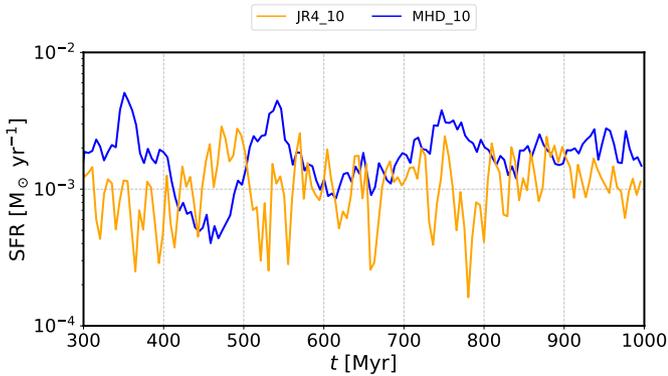

**Figure A4.** The star formation rate (SFR) for JR4_10 and MHD_10 over the steady state period. Model JR4_10 has more variability in its SFR but on average is almost the same at $1.21 \times 10^{-3}$ M$_\odot$ yr$^{-1}$.

Including the sink particles does not result in a large scale change to the molecular gas within the models. These values are an upper limit on the mass fractions and depletion times. A more detailed analysis of including sink mass for the hydrodynamical models can be found in the Appendix of paper 1.

This paper has been typeset from a T$_{\rm E}$X/L$^{\rm A}$T$_{\rm E}$X file prepared by the author.

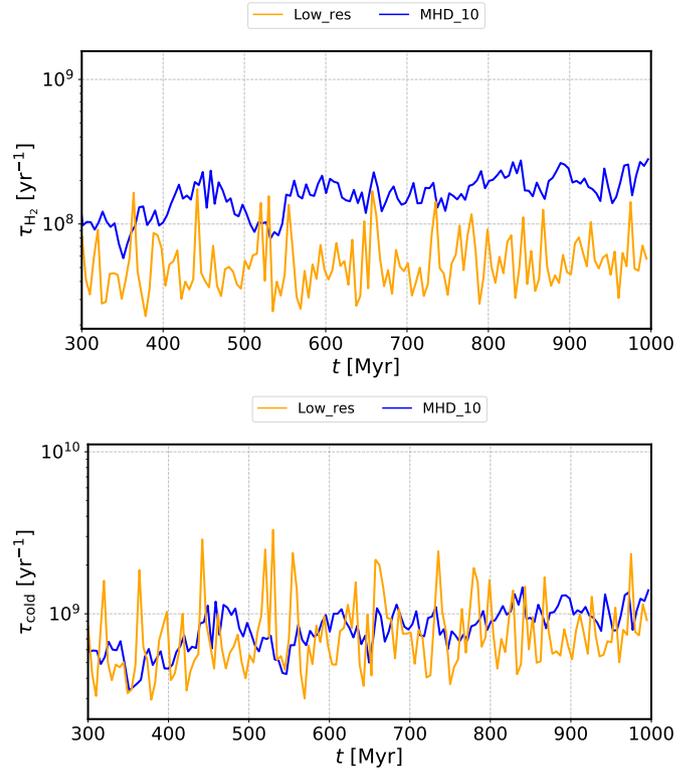

**Figure A5.** The molecular gas depletion times ($\tau_{\rm H_2}$) (top) and cold gas depletion times ($\tau_{\rm cold}$) (bottom) for models JR4_10 and MHD_10. The lower resolution JR4_10 has a lower $\tau_{\rm H_2}$ than MHD_10 which arises due to not fully resolving the dense gas, the $\tau_{\rm cold}$ for both models are roughly the same.

| Model | $\frac{M_{\rm H_2}}{M_{\rm HI+H_2}}$ | $\frac{M_{\rm H_2+sink}}{M_{\rm HI+H_2}}$ | H$_2$ | | H$_2$ + sink | |
|---|---|---|---|---|---|---|
| | | | $\tau_{\rm dep}$ (Myr) | $\sigma_{\tau_{\rm dep}}$ (Myr) | $\tau_{\rm dep}$ (Myr) | $\sigma_{\tau_{\rm dep}}$ (Myr) |
| MHD_01 | 0.24% | 0.78% | 73.3 | 36.5 | 238 | 94.1 |
| MHD_10 | 0.43% | 0.85% | 163 | 61.6 | 321 | 104 |
| MHD_SAT | 0.56% | 0.89% | 263 | 83.2 | 421 | 126 |

**Table B1.** Mass fractions and depletion times $\tau_{\rm dep}$ for H$_2$ and H$_2$ including fully molecular sinks with their standard deviations, $\sigma_{\tau_{\rm dep}}$ for the MHD models. MHD_01 shows an increase of a factor of $\sim 3$ in both mass fraction and depletion times when sinks are included. MHD_10 and MHD_SAT increase by a factor of $\sim 2$ and $\sim 1.6$ respectively.

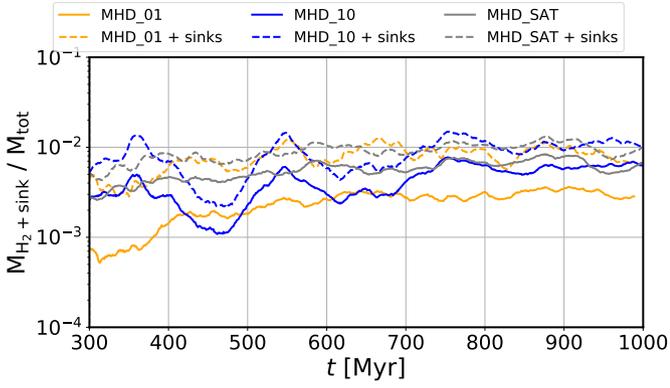

**Figure B1.** Molecular gas mass fraction of of the MHD models including fully molecular sinks (dashed line) and without including the sink mass (solid lines). In MHD_10 and MHD_SAT there is a small increase in the mass fraction, with a larger increase in model MHD_01.

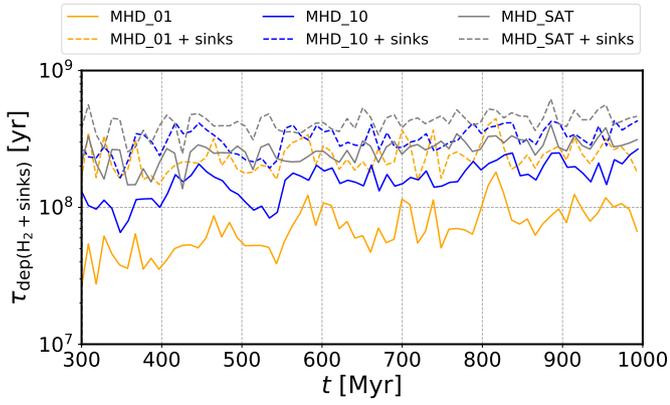

**Figure B2.** Molecular gas depletion time ($\tau_{H_2}$) of of the MHD models including fully molecular sinks (dashed line) and without including the sink mass (solid lines). As with the mass fraction we see a small increase in MHD_10 and MHD_SAT, with a slightly larger increase in model MHD_01.